\documentclass[11pt]{article}

\usepackage{verbatim}
\usepackage{amsmath,amscd}
\usepackage{amsfonts}
\usepackage{amssymb}
\usepackage{bbm}
\usepackage{latexsym}
\usepackage{graphicx}
\usepackage{lscape}
\usepackage{epsfig}
\usepackage{color}
\usepackage{tikz}
\usepackage{multirow}
\usepackage{afterpage}
\usepackage{amscd}
\usepackage{cite}
\usepackage{tabu}

\usepackage{cancel}
\usepackage{lmodern}

\usepackage{hyperref}
\usepackage{longtable}
\usepackage[font=normalsize]{caption}
\usepackage{microtype}

\newcommand{\enums}[1]{\begin{enumerate} #1 \end{enumerate}}
\newcommand{\equ}[1]{\begin{gather} #1 \end{gather}}
\newcommand{\equa}[1]{\begin{align} #1 \end{align}}
\newcommand{\non}{\nonumber}
\newcommand{\sfrac}[2]{\mbox{$\frac{#1}{#2}$}}

\newcommand{\arry}[2]{\begin{array}{#1} #2 \end{array}}
\newcommand{\mtrx}[1]{\begin{matrix} #1 \end{matrix}}
\newcommand{\pmtrx}[1]{\begin{pmatrix} #1 \end{pmatrix}}
\newcommand{\brkt}[2]{\bigl[ ^{#1}_{#2} \bigr]}

\newcommand{\Z}[1]{\ensuremath{\mathbbm{Z}_{#1}}} 
\newcommand{\Intr}{\ensuremath{\mathbbm{Z}}}
\newcommand{\Real}{\ensuremath{\mathbbm{R}}}
\newcommand{\Ratl}{\ensuremath{\mathbbm{Q}}}
\newcommand{\Cplx}{\ensuremath{\mathbbm{C}}}

\newcommand{\I}{\mathrm{i}}
\renewcommand{\d}{\mathrm{d}}

\renewcommand{\gg}{\gamma}

\newcommand{\ga}{\alpha}
\newcommand{\gb}{\beta}
\newcommand{\gd}{\delta}
\newcommand{\gc}{\chi}
\newcommand{\gm}{\mu}
\newcommand{\gn}{\nu}
\newcommand{\get}{\eta}
\newcommand{\gk}{\kappa}
\newcommand{\gl}{\lambda}
\newcommand{\gr}{\rho}
\newcommand{\gth}{\theta}
\newcommand{\gs}{\sigma}
\newcommand{\gt}{\tau}
\newcommand{\go}{\omega}
\newcommand{\gps}{\psi}
\newcommand{\gch}{\chi}

\newcommand{\bgt}{{\bar\tau}}
\newcommand{\bgg}{{\bar\gamma}}
\newcommand{\bgs}{{\bar\sigma}}

\newcommand{\gL}{\Lambda}
\newcommand{\gG}{\Gamma}
\newcommand{\gS}{\Sigma}
\newcommand{\tgG}{{\tilde \Gamma}}

\newcommand{\cF}{{\cal F}}
\newcommand{\cN}{{\cal N}}

\newcommand{\rep}[1]{\mathbf{#1}}
\newcommand{\crep}[1]{\overline{\rep{#1}}}

\newcommand{\E}[1]{\ensuremath{\mathrm{E_{#1}}}}
\newcommand{\U}[1]{\ensuremath{\mathrm{U(#1)}}}
\newcommand{\SU}[1]{\ensuremath{\mathrm{SU(#1)}}}
\newcommand{\SO}[1]{\ensuremath{\mathrm{SO(#1)}}}
\newcommand{\Spin}[1]{\ensuremath{\mathrm{Spin(#1)}}}
\newcommand{\ch}[2]{{{\gch}_{#1}{\left(#2\right)}}}

\newcommand{\Ra}{\Rightarrow}
\newcommand{\ra}{\rightarrow}
\newcommand{\Id}{\mathbbm{1}}

\usepackage[colorinlistoftodos]{todonotes}

\newtheorem{conjecture}{Conjecture}

\begin{document}

\thispagestyle{empty}

\begin{flushright}
TUM-HEP 1104/17 \\
\end{flushright}
\vskip .2 cm
\begin{center}
{\Large {\bf Tension Between a Vanishing Cosmological Constant and Non-Supersymmetric Heterotic Orbifolds} } \\[0pt]

\bigskip
\bigskip {\large
{\bf Stefan Groot Nibbelink$^{a,}$}\footnote{E-mail: groos@hr.nl},
{\bf Orestis Loukas$^{b,}$}\footnote{E-mail: orestis.loukas@cern.ch},
{\bf Andreas M\"utter$^{c,}$}\footnote{E-mail: andreas.muetter@tum.de},
{\bf Erik Parr$^{c,}$}\footnote{E-mail: erik.parr@tum.de},
{\bf Patrick K.S. Vaudrevange$^{c,}$}\footnote{E-mail: patrick.vaudrevange@tum.de},
\bigskip }\\[0pt]
\vspace{0.23cm}
${^a}$ {\it School of Engineering and Applied Sciences, Rotterdam University of Applied Sciences, \\
G.J.\ de Jonghweg 4 - 6, 3015 GG Rotterdam, Netherlands}\\[1ex]
${^b}$ {\it Albert Einstein Center for Fundamental Physics,  Institute for Theoretical Physics,\\ University of Bern,
Sidlerstrasse 5, ch-3012 Bern, Switzerland}\\[1ex]
$^c$ {\it Physik Department T75, Technische Universit\"at M\"unchen, \\
James-Franck-Stra\ss e, 85748 Garching, Germany\\[1ex]
}

\bigskip
\end{center}

\subsection*{\centering Abstract}

We investigate under which conditions the cosmological constant vanishes perturbatively at the 
one-loop level for heterotic strings on non-supersymmetric toroidal orbifolds. To obtain 
model-independent results, which do not rely on the gauge embedding details, we require that 
the right-moving fermionic partition function vanishes identically in every orbifold sector.
This means that each sector preserves at least one, but not always the same Killing spinor. 
The existence of such Killing spinors is related to the representation theory of finite groups, 
i.e.\ of the point group that underlies the orbifold.
However, by going through all inequivalent (Abelian and non-Abelian) point groups of six-dimensional toroidal orbifolds 
we show that this is never possible: For any non-supersymmetric orbifold there is always 
(at least) one sector, that does not admit any Killing spinor. The underlying mathematical reason 
for this no-go result is formulated in a conjecture, which we have tested by going through an even 
larger number of finite groups. This conjecture could be applied to situations beyond symmetric 
toroidal orbifolds, like asymmetric orbifolds.

\newpage
\setcounter{page}{1}
\setcounter{footnote}{0}
%
%


\section{Introduction}


The question why the cosmological constant is very small compared to any other scale in physics, 
yet non-zero (thereby driving the current expansion of the universe), is possibly one of the most 
challenging ones in modern physics. Determining it from first principles presumably involves a 
detailed understanding of quantum gravity. String theory is often suggested as a theory for quantum 
gravity and within that framework the cosmological constant can be computed, at least in principle. 
Furthermore, in heterotic  string theory a non-vanishing cosmological constant is associated 
with a non-vanishing dilaton tadpole, which possibly signifies that one is not expanding the theory 
around a stable point~\cite{Fischler:1986ci,Fischler:1986tb}. Consequently, a vanishing 
cosmological constant (at least at one loop) is also instrumental to avoid the dilaton tadpole.


The smallness of the cosmological constant could be taken as a hint for an underlying symmetry: The 
symmetry should be such that it forces the cosmological constant to vanish, at least at tree- and 
one-loop level in string perturbation theory. In such a setting, a small breaking of this symmetry 
would generate a small vacuum energy in the theory, hence introducing a small value for the cosmological constant. 
Supersymmetry seemed to be a promising candidate, as it is known that the cosmological constant (once it vanishes at tree-level) is identically zero in supersymmetric field theories to all orders in ordinary perturbation theory, due to non-renormalization theorems. There are strong indications \cite{green1988superstring,dine1988string} that target-space supersymmetry remains unbroken to all orders in string perturbation theory and thus, similar non-renormalization theorems also apply in the string-theoretic setup ensuring a perturbative vanishing of the vacuum energy and the predicted cosmological constant. However, supersymmetry has to be eventually broken and the relevant scale of 
breaking must be much larger than the weak scale (also due to experimental bounds from the LHC).  
This re-introduces a huge fine-tuning problem for the cosmological constant making scenarios  with soft breaking of supersymmetry less phenomenologically viable. 
Therefore, we ask whether there exist non-supersymmetric string models 
which have new (stringy) mechanisms to predict a vanishing cosmological constant.


Contrary to what is sometimes claimed, string theory does not require target-space supersymmetry. 
Indeed, the first example of a non-supersymmetric but otherwise consistent string theory is the 
SO(16)$\times$SO(16) string~\cite{Dixon:1986iz,Dixon:1986jc,AlvarezGaume:1986jb}. Torus 
compactifications with Wilson lines were first considered in~\cite{Nair:1986zn,Ginsparg:1986wr} and 
described in a covariant lattice approach~\cite{Lerche:1986ae,Lerche:1986cx}. Orbifold 
compactifications of this theory have been investigated in~\cite{Taylor:1987uv,Toon:1990ij,Sasada:1995wq,Font:2002pq} and described in the free-fermionic formulation~\cite{Dienes:1994np,Blum:1997cs,Shiu:1998he,Dienes:2006ut,Faraggi:2007tj}. Such constructions can lead to models that possess low-energy spectra quite close to the Standard Model of Particle Physics~\cite{Blaszczyk:2014qoa,Blaszczyk:2015zta,Nibbelink:2015vha,Nibbelink:2015ena,Abel:2015oxa,Ashfaque:2015vta} 
(which might possess fermionic symmetries even though being non-supersymmetric~\cite{Nibbelink:2016lzi}).
Moreover, for some of such models more detailed phenomenological quantities, like gauge threshold corrections, were investigated~\cite{Angelantonj:2014dia,Florakis:2015txa}. 
Also the value of the cosmological constant can be computed explicitly for such non-supersymmetric 
string models from the string partition function. However, as far as we are aware, there exists no 
explicit non-supersymmetric heterotic construction in the literature with identically 
vanishing cosmological constant.


This is in contrast to the status for the type II string in which non-supersymmetric models were 
obtained with one-loop vanishing cosmological constant on asymmetric orbifolds in 
ref.~\cite{Kachru:1998hd,Shiu:1998he,Kachru:1998pg,Blumenhagen:1998uf,Satoh:2015nlc}. 
The asymmetric nature of these constructions is crucial for the cosmological constant to vanish at 
one loop. In all sectors either the left-moving partition function vanishes, while the right-moving 
counterpart does not, or the other way around. Nevertheless, the initial hope that this feature 
persists at two-loop order and beyond was questioned in~\cite{Iengo:1999sm} where it was argued 
that the relevant integrand does not vanish at two loops.


The main goal of the current paper is to investigate whether it is possible to construct 
non-supersymmetric heterotic string models on toroidal orbifolds in which the cosmological constant 
vanishes identically at the one-loop level in a rather model independent way, i.e.\ without relying 
on the details of the gauge embedding. In fact, we will show that this is impossible for 
symmetric toroidal orbifolds and even extremely unlikely for asymmetric ones. To this end we have 
structured the paper as follows:

\subsection*{Overview}


We begin in section~\ref{sec:PartitionFunction} with a review of the necessary ingredients for the 
computation of the one-loop cosmological constant in our setup: the orbifold point and space groups 
and the structure of the one-loop string partition function in terms of (twisted-)sectors 
characterized by constructing and projecting space group elements. We identify three levels at 
which the one-loop partition function and, hence, the cosmological constant can be forced to 
vanish. To choose the most model-independent option, we require that the one-loop partition 
function vanishes in each (twisted-)sector individually. We subsequently argue that this only 
happens if the right-moving fermionic partition function vanishes for all these sectors separately. 
Hence, any commuting pair of constructing and projecting space group elements need to have at least 
one Killing spinor in common. In such a setting, each (twisted-)sector of the partition function 
would have a boson-fermion degeneracy similar to supersymmetry, even though the model as a whole is 
non-supersymmetric. Precisely, these degeneracies would force the one-loop vacuum-to-vacuum 
amplitude and hence the leading contribution to the cosmological constant to vanish identically.


In section~\ref{sec:ExplicitNoGo} we show that apart from supersymmetry preserving orbifolds, all 
toroidal orbifolds posses some space group elements that do not admit any Killing spinors. 
Consequently, for all non-supersymmetric toroidal orbifolds the right-moving fermionic partition 
function does not vanish in each sector individually. We prove this by explicitly constructing all 
possible spin embeddings of all point groups associated to six-dimensional symmetric toroidal 
orbifolds, relying on their classification~\cite{Opgenorth:1998ab,Plesken:2000ab}.


In section~\ref{sec:FiniteGroupNoGo} we provide an alternative, more sophisticated, proof by making 
use of finite group theory. This leads us to formulate a conjecture about the non-existence of 
finite groups with a four-dimensional representation possessing certain mathematical properties. We 
confirmed this conjecture for all $\mathcal{O}(100,000)$ finite groups of order smaller or equal 
500. Finally, based on this conjecture we present some arguments that our no-go result extends even 
beyond symmetric toroidal orbifolds. 


Since the presentation in this paper is rather abstract, we illustrate various aspects 
using a number of examples based on the non-Abelian group $Q_8$. In particular, important double 
cover ambiguities that we needed to resolve in the general proofs are exemplified by various 
$\Z{2}\times\Z{2}$ and $Q_8$ orbifolds.


In section~\ref{sec:conclusion} we state our main conclusions and put them in perspective by making 
some comparisons with the existing literature. Moreover, we suggest future 
extensions and applications of the results of this paper. 


We have collected technical details in a number of appendices: Appendix~\ref{App:AppRiemannId} 
describes some elements of one-loop fermionic partition functions. Furthermore, we derive 
generalized Riemann identities which provide conditions for the building blocks of one-loop 
partition functions to vanish identically. Since the spinorial embedding with associated double 
cover ambiguities is of the utmost importance to guarantee that we considered all possible cases in 
our nonexistence proofs, we reviewed the relevant SO(6), Spin(6) and SU(4) representation theory in 
appendix~\ref{App:Spin}. Some important aspects of finite groups and their representations used in 
section~\ref{sec:FiniteGroupNoGo} are collected in appendix~\ref{app:FiniteGroup}.

\section{Partition Functions of Heterotic Toroidal Orbifolds}
\label{sec:PartitionFunction} 

\subsection{One-Loop Cosmological Constant in Heterotic String Theories}
\label{sec:cosmoconstant} 


At the one-loop level the four-dimensional cosmological constant (or Casimir energy density) 
$\gL$  of the heterotic string is proportional to  
\equ{ \label{GeneralCosmoC}
\gL ~\sim~ 
\int_\cF \frac{\d^2\gt}{\gt_2^2}\, \mathcal{Z}_\text{full}(\gt,\bgt)~.
}
%
The modular integral is over the fundamental domain $\cF$ of the Euclidean worldsheet torus 
characterized by the complex Teichm\"uller parameter $\gt$ with 
$\cF=\lbrace \gt = \gt_1 + \I \gt_2\,|\,-\frac{1}{2}\leq \gt_1 \leq \frac{1}{2}, \gt_2 > 0, |\gt|>1\rbrace$. 
This modular parameter $\gt$ is defined modulo modular transformations. The factor $1/\gt_2^2$ has 
been introduced in eq.~\eqref{GeneralCosmoC} in order to obtain a modular invariant measure. Consequently, 
the full one-loop partition function $\mathcal{Z}_\text{full}$ has to be modular invariant by itself.

The full one-loop partition function can be factorized into a four-dimensional non-compact part 
$\mathcal{Z}_\text{Mink.}$ and an internal part $\mathcal{Z}_\text{int.}$, i.e.
\equ{ \label{FacFullPI} 
\mathcal{Z}_\text{full}(\gt,\bgt) ~=~ \mathcal{Z}_\text{Mink.}(\gt,\bgt) \, \mathcal{Z}_\text{int.}(\gt,\bgt)~.
}
The non-compact part corresponds to the coordinate fields $x=(x^\gm)$, $\gm =0,1,2,3$ living in 
the four-dimensional Minkowski space $\Real^{1,3}$. The corresponding one-loop partition function 
in light-cone gauge reads
\equ{ \label{noncompactPI}
\mathcal{Z}_\text{Mink.}(\gt,\bgt)  ~=~ \frac 1{\gt_2} \Big| \frac1{\get^{2}(\gt)} \Big|^2~, 
}
where $\get(\gt)$ is the Dedekind function.

\subsection{Space and Point Groups of Toroidal Orbifolds}
\label{sec:orbifoldgeometry}


A six-dimensional toroidal orbifold~\cite{Dixon:1985jw,Dixon:1986jc} can be constructed in two 
steps: First, one chooses a six-dimensional lattice $\gG$ spanned by a vielbein 
$e \in \text{GL}(6;\Real)$, such that a general vector in the lattice $\gG$ is uniquely 
parametrized by $m \in \Z{}^6$ as $e\,m \in \gG$. Next, one defines a six-dimensional torus 
$T^6 = \Real^6/\gG$ with a metric 
\equ{ \label{TorusMetric}
G=e^Te~,
}
where the quotient space $\Real^6/\gG$ is given by identifying those points in $\Real^6$ that 
differ by any lattice vector from $\gG$. In a second step, one chooses an abstract finite group 
$\mathbf{P}$ and a six-dimensional representation $D_\rep{v}: \mathbf{P} \ra \SO{6}$. Then, we 
introduce the so-called geometrical point group as 
$\mathbf{P}_\rep{v} = D_\rep{v}(\mathbf{P}) \subset \SO{6}$, which is defined as a finite 
matrix group of discrete lattice automorphisms of the torus lattice $\gG$. Given that 
$D_\rep{v}(\gth) \in \mathbf{P}_\rep{v}$ is an automorphism of the torus lattice for each 
$\gth \in \mathbf{P}$, one finds 
\begin{equation}\label{eqn:chrystallographicaction}
D_\rep{v}(\gth)\,e ~=~ e\, \widehat{D}_\rep{v}(\gth) \quad\text{with}\quad \widehat{D}_\rep{v}(\gth) \in \text{GL}(6;\Z{})~,
\end{equation}
where $\widehat{D}_\rep{v}(\gth)$ is the so-called twist in the lattice basis. (We use the hatted 
notation to emphasize that $\widehat{D}_\rep{v}(\gth)$ is an integral matrix.) A toroidal orbifold 
is now defined by identifying those points on $T^6$ that are mapped onto each other by elements of 
$\mathbf{P}_\rep{v}$.


The translational and rotational actions of the lattice group $\gG$ and the  point group 
$\mathbf{P}$ can be combined to build elements of the form $(\gth,e\,m)\in\mathbf{S}$ of the 
so-called space group $\mathbf{S}$. An element $g =(\gth, e\, m) \in \mathbf{S}$ acts on the 
six-dimensional torus as 
\equ{
g \circ X ~=~ D_\rep{v}(\gth)\, X + e\,m~. 
}
The smallest positive integer $N_\theta$ for which $\theta^{N_\theta} = \Id$ is called the 
order of the point group element $\theta$. For space group elements $g', g \in \mathbf{S}$ we find 
the following product rule: 
\equ{
g'\, g ~=~ \left(\gth', e\,m'\right)\,\left(\gth, e\,m\right) ~=~ \left(\gth'\, \gth, e\, m' + D_\rep{v}(\gth')\, e\, m\right)~. 
}
%


The space group may also include elements called roto-translations. A roto-translation combines the 
action of a point group element $\gth\in \mathbf{P}$ of order $N_\gth$ with a simultaneous 
translation by a fractional lattice vector, i.e.\ $(\gth, e\, \gm) \in \mathbf{S}$ where 
$\gm \in \Ratl^6$ such that $(\gth, e\, \gm)^{N_\gth} = (\Id, e\, m)$ with $m\in\Z{}^6$ using 
$\gth^{N_\gth} = \Id$. 

In summary, a toroidal orbifold is geometrically defined by the choice of a geometrical point group 
$\mathbf{P}_\rep{v}$ and its extension to a space group $\mathbf{S}$. However, to define string 
theory (or target-space field theory with spinors) on toroidal orbifolds we need to specify the 
action of the point group on target-space spinors, as we discuss next.

\subsection{Point Group Action on Target-Space Spinors}
\label{sec:DoubleCover} 


In the heterotic orbifold literature, the groups $\mathbf{P}_\rep{v}$ and $\mathbf{P}$ are often 
implicitly identified and interpreted in a purely geometrical fashion. However, for string theory 
we also need to specify the action of $\mathbf{P}$ on target-space spinors, which is defined by the 
eight-dimensional (reducible) spinor representation $D_\rep{s}: \mathbf{P} \ra \text{Spin}(6)$, 
see appendix~\ref{App:Spin}. In fact, we assume that the spinor representation $D_\rep{s}$ is 
faithful such that the corresponding matrix group 
$\mathbf{P}_\rep{s} = D_\rep{s}(\mathbf{P}) \subset \text{Spin}(6)$ is isomorphic to the abstract 
point group $\mathbf{P}$. In other words, we define the abstract point group $\mathbf{P}$ by its 
action on spinors and allow for the case $\mathbf{P}_\rep{v} \subseteq \mathbf{P}$. Consequently, we 
distinguish between the abstract point group $\mathbf{P}$ with elements denoted by 
$\gth \in \mathbf{P}$ and two representations of $\mathbf{P}$: the six-dimensional (possibly 
non-faithful) representation $D_\rep{v}$ and the (faithful) eight-dimensional spinorial 
representation $D_\rep{s}$.

The distinction between $\mathbf{P}_\rep{v}$ and $\mathbf{P}$ is important, since the spin group 
$\text{Spin}(6)$ is the double cover of the orthogonal group $\SO{6}$: any element 
$D_\rep{v}(\gth)$ of the geometrical point group $\mathbf{P}_\rep{v}$ has two representatives in 
the $\text{Spin}(6)$ group as $\pm D_\rep{s}(\gth)$, see appendix~\ref{App:Spin} and 
eq.~\eqref{DoubleCover} therein. Hence, if the geometrical point group $\mathbf{P}_\rep{v}$ has a 
maximal set of $K$ generators, then in principle we could have up to $2^K$ possibilities for the 
action of the point group on spinors. 
If the order $N_\gth$ of a generator $\gth$ is odd, then the 
order of the corresponding spin generator can be doubled depending on the choice of sign: 
Since $D_\rep{s}$ is a faithful representation of $\mathbf{P}$, it follows that $D_\rep{s}(\gth)$ 
has the same order $N_\gth$ as $\gth\in\mathbf{P}$ and consequently $-D_\rep{s}(\gth)$ is of order 
$2N_\gth$. 

The finite group $\mathbf{P}_\rep{v}$ is defined by a so-called presentation, i.e.~by a number of 
defining relations among its generators. These defining relations can always be cast in the form of 
a product of $\mathbf{P}_\rep{v}$ elements that equals the identity. Since $\Spin{6}$ is the double 
cover of $\SO{6}$, the defining relations for the corresponding $\mathbf{P}_\rep{s}$ can take two 
forms, i.e.
\equ{ \label{ModificationsDefRelations}
D_\rep{v}(\gth_1) \cdots D_\rep{v}(\gth_n) ~=~ \Id_6
\qquad\Ra\qquad 
D_\rep{s}(\gth_1) \cdots D_\rep{s}(\gth_n) ~=~ \pm \Id_8~. 
}
Indeed, the relation between the vector and spinor representations eq.~\eqref{DoubleCover} implies 
by Schur's lemma (derived around eq.~\eqref{CliffordSchurLemma}) that each defining relation in the 
spinor representation is necessarily equal to the identity matrix up to a sign. Now, if all 
defining relations for $\mathbf{P}_\rep{s}$ are the same as those of $\mathbf{P}_\rep{v}$, then 
$\mathbf{P}_\rep{s}$ and $\mathbf{P}_\rep{v}$ are isomorphic. Otherwise, at least one defining 
relation of $\mathbf{P}_\rep{s}$ is equal to $-\Id_8$ and hence necessarily also 
$-\Id_8 \in \mathbf{P}_\rep{s}$. The latter element describes a ten-dimensional orbifold sector 
with a non-trivial action on spinors.

Of course, the other way around there is no ambiguity: If the spinor representation $D_\rep{s}$ has 
been specified, then the vector representation $D_\rep{v}$ is uniquely determined, see 
eq.~\eqref{UniqueSOrep}. For toroidal orbifolds that preserve supersymmetry these ambiguities 
concerning the double cover can be ignored safely, as there is a unique supersymmetric assignment 
for both the vector and the spinor representations. However, given that the current project is 
about non-supersymmetric models, these different choices, in fact, correspond to different models, 
with possibly different cosmological constants.


The subtlety with the double cover can be illustrated nicely with the concept of the local twist 
vector. By a basis change any element $D_\rep{v}(\gth) \in \SO{6}$ (associated to 
$g=\left(\gth, e\,m\right) \in \mathbf{S}$) can be block-diagonalized such that it acts as a 
rotation in the three orthogonal planes given by $(X^1,X^2)$, $(X^3,X^4)$ and $(X^5,X^6)$. Then, 
using complex coordinates $Z^a = X^{2a-1} + \I X^{2a}$ for $a=1,2,3$ the rotation in each 
$(X^{2a-1},X^{2a})$-plane is represented by a simple phase transformation, i.e.\ 
$Z^a \mapsto \exp{\left(2\pi \I\, v_g^a\right)}\,Z^a$. Using eq.~\eqref{SigmaBasis} we can cast the 
vector and spinor representations, $D_\rep{v}(\gth)$ and $D_\rep{s}(\gth)$, associated to the point 
group element $\gth \in \mathbf{P}$ of $g$ in the following forms  
\begin{subequations}\label{eq:localTwistBasis} 
\begin{eqnarray}
D_\rep{v}(\gth) & = & \pmtrx{ 
e^{2\pi \I\, v_g^1} & 0 & 0 \\[1ex]  
0 & e^{2\pi \I\, v_g^2} & 0 \\[1ex]  
0 & 0 & e^{2\pi \I\, v_g^3} }~, \label{eq:localTwistBasisA} \\[2ex] 
D_\rep{s}(\gth) & = & 
e^{2\pi \I\, v_g^1\, \frac 12\gs_3} \otimes e^{2\pi \I\, v_g^2\, \frac 12\gs_3} \otimes e^{2\pi \I\, v_g^3\, \frac12 \gs_3}~.
\label{eq:localTwistBasisB} 
\end{eqnarray}
\end{subequations} 
They are both parametrized by the same local twist vector $v_g=\big( 0, v_g^1,v_g^2,v_g^3\big)$, 
where we introduced an additional zero-entry as $a=0$ component of $v_g$ for later use. Note that 
as far as the vector representation $D_\rep{v}(\gth)$ is concerned each of the components of the 
local twist vector has a periodicity $v_g^a \sim v_g^a +1$ that leaves $D_\rep{v}(\gth)$ invariant. 
However, the spinor representation $D_\rep{s}$ is only invariant when two of such identifications 
are combined, i.e.\ the local twist vector $v_g$ is defined modulo vectors from the root lattice of 
$\SO{8}$. In particular, the twist vectors $\big( 0, v_g^1,v_g^2,v_g^3\big)$ and 
$\big( 0, v_g^1,v_g^2,v_g^3+1\big)$ give rise to the same $\SO{6}$-matrix $D_\rep{v}(\gth)$ but to 
two different $\text{Spin}(6)$-matrices $D_\rep{s}(\gth)$ and $-D_\rep{s}(\gth)$, respectively. 
Hence, this ambiguity in defining $v_g$ distinguishes between the two Spin$(6)$ representatives in 
the double cover of the $\SO{6}$-matrix $D_\rep{v}(\gth)$.

The non-trivial embedding of the geometrical point group action into spinor space can be implemented 
to understand supersymmetry breaking using roto-translations. In most cases, a roto-translation 
$(\gth, e\, \gm)$ contains some proper rotation, i.e.\ $D_\rep{v}(\gth) \neq \Id_6$. Scherk-Schwarz 
supersymmetry breaking can be thought of as a special case of such roto-translations: Even though 
the twist action is trivial, i.e.\ $D_\rep{v}(\gth)=\Id_6$, the action on the spinors is not, 
$D_\rep{s}(\gth)=-\Id_8$. At the group-theoretic level, $\gth$ generates a $\Z{2}$, whose 
geometrical embedding is trivial, $\mathbf{P}_\rep{v}=\{\Id_6\}$, while $\mathbf{P}_\rep{s}$ is 
isomorphic to the abstract point group $\mathbf{P} = \Z{2}$. In other words, there exists a 
ten-dimensional orbifold sector where supersymmetry is broken by the non-trivial action of 
$\mathbf{P}_\rep{s}$. (An equivalent description of Scherk-Schwarz supersymmetry breaking uses a 
fractional translation with a factor $(-1)^{F_\text{R}}$ in the partition function, where 
$F_\text{R}$ is the right-moving fermion number~\cite{FERRARA198975}.)

\subsection{Structure of Orbifold Partition Functions}
\label{sec:OrbifoldPartitionFct}


In heterotic string theory, the six-dimensional toroidal orbifold geometry gives rise to 
additional boundary conditions for the corresponding coordinate fields $X = (X^i)$, with 
$i=1,\ldots,6$. In detail, strings on toroidal orbifolds can close up to the action of an element 
$g  = (\gth, e\,m)\in \mathbf{S}$,
\begin{equation}\label{eqn:boundaryconditiong}
X(\sigma_1+1,\sigma_2) ~=~ g \circ X(\sigma_1,\sigma_2) ~=~ D_\rep{v}(\gth)\,X(\sigma_1,\sigma_2) + e\,m~.
\end{equation}
We call $g$ the constructing element of the string. For $D_\rep{v}(\gth) \neq \Id_6$ these boundary 
conditions give rise to so-called twisted strings.


In addition to the fields corresponding to the geometry of the orbifold, there are two other types 
of worldsheet fields to complete the heterotic theory: First of all, the theory contains eight 
real fermionic partners of the coordinate fields $(x^\gm, X^i)$ $\gm=2,3$ in light-cone gauge and 
$i=1,\ldots 6$. However, it is often more convenient to group them as four complex fermions 
$\gps_\text{R} = (\gps_\text{R}^a)$ for $a=0,1,2,3$. The index $a=0$ corresponds to the two 
non-compact directions transversal to the light-cone and $a=1,2,3$ to the three complexified torus 
directions. Secondly, there are sixteen additional left-moving coordinates 
$Y_\text{L} = (Y^I_\text{L})$ with $I=1,\ldots, 16$ that take values on a sixteen dimensional torus. 
(Equivalently, the latter degrees of freedom are also often described by 16 left-moving complex 
fermions $\gl_\text{L} = (\gl_\text{L}^I)$.)


Since the one-loop partition function corresponds to a worldsheet with the topology of a torus, the 
closed string eq.~\eqref{eqn:boundaryconditiong} is subject to a second boundary condition
\equ{\label{eqn:boundaryconditionh}
X(\sigma_1 + \tau_1, \sigma_2 + \tau_2) ~=~ h  \circ X(\sigma_1, \sigma_2)\;,
}
with $h \in \mathbf{S}$. To make the boundary condition eq.~\eqref{eqn:boundaryconditiong} 
compatible with eq.~\eqref{eqn:boundaryconditionh}, the associated space group elements 
$g, h \in \mathbf{S}$ have to commute, i.e.\ $h\, g = g\, h$. 
Then, the orbifold partition function can be organized as a sum over sectors characterized by commuting $g$ and $h$, namely
\equ{ \label{IntTwistedPI}
\mathcal{Z}_\text{int.}(\tau,\overline{\tau}) ~=~ \frac 1{|\mathbf{P}|} \sum_{[g,h]=0} \mathcal{Z}_\text{int.}\brkt{g}{h}(\tau,\overline{\tau})~.
}
Even though the full point group $\mathbf{P}$ can be non-Abelian, the point group elements 
associated to the commuting elements $g$ and $h$ can be diagonalized simultaneously. Using 
eq.~\eqref{eq:localTwistBasis} this yields the local twist vectors $v_g$ and $v_h$. Furthermore, 
the internal partition function of each $(g,h)$-twisted sector can be factorized as follows 
\equ{ \label{FacIntTwistedPI} 
\mathcal{Z}_\text{int.}\brkt{g}{h}(\tau,\overline{\tau})~=~ 
\overline{\mathcal{Z}_\gps\brkt{g}{h}(\tau)}\, 
\mathcal{Z}_X\brkt{g}{h}(\gt,\bgt) \, 
\mathcal{Z}_Y\brkt{g}{h}(\tau)\;, 
}
associated with the contributing worldsheet fields $\gps_\text{R}$, $X$ and $Y_\text{L}$. The 
explicit forms of the last contribution $\mathcal{Z}_Y$ in eq.~\eqref{FacIntTwistedPI} is highly 
model-dependent via the choice of the gauge embedding. Such choices are of course constrained by 
modular invariance\footnote{Since our results mostly only rely on the right-moving sector, we do 
not dwell on modular invariance here and use that the standard embedding always produces 
a consistent choice.}. The partition function associated with the 
right-moving fermions $\gps_\text{R}$ encodes all the information about target-space supersymmetry. 
This part only depends on the space group $\mathbf{S}$, but not on the particular gauge embedding 
chosen. Furthermore, $\mathcal{Z}_\gps\brkt{g}{h}(\tau)$ depends only on the local twist 
vectors $v_g$ and $v_{h}$ (defined above) corresponding to 
the commuting space group elements $g$ and $h$, respectively, and not on the torus 
lattice, i.e.
\begin{equation}
\mathcal{Z}_\gps\brkt{g}{h}(\tau) ~=~ \mathcal{Z}_4\brkt{v_g}{v_{h}}(\gt)~, 
\end{equation}
see eq.~\eqref{RMpartitionFunctionOfPsi} in appendix~\ref{app:RMPartitionFunction}. Hence, it is 
the same for huge families of orbifold models.

\subsection{A Vanishing One-Loop Cosmological Constant}
\label{sec:vanishingcc}

From the discussion above it is clear that the one-loop cosmological constant~\eqref{GeneralCosmoC} 
is in general a complicated function of the moduli of the compactification and its value 
sensitively depends on the type of compactification manifolds in question. Hence, instead of 
trying to determine its expression in general, we ask the question under which conditions it 
vanishes identically.

Based on the structure of the full partition function eqs.~\eqref{IntTwistedPI} 
and~\eqref{FacIntTwistedPI}, we realize that there are various ways the resulting cosmological 
constant may vanish at one loop. Let us list the various logical options in a nested way and 
briefly comment on each possibility: 
\enums{
\item[{\bf 1.}] {\bf The full partition function $\boldsymbol{\mathcal{Z}_\text{full}}$, i.e.\ the 
integrand of eq.~$\boldsymbol{\eqref{GeneralCosmoC}}$, is non-vanishing, but nevertheless the cosmological 
constant, i.e.\ the integral, is zero.} 
\\[1ex] 
This option may be realized by a generalization of Atkin-Lehner symmetry (first introduced in this context by \cite{MOORE1987139}, see also \cite{PhysRevD.42.2004}) and hence requires a 
detailed understanding of the modular properties of the full partition function. 
\item[{\bf 2.}] {\bf The full partition function $\boldsymbol{\mathcal{Z}_\text{full}}$ vanishes, 
but some of the partition functions $\boldsymbol{\mathcal{Z}_\text{int.}\brkt{g}{h}}$ associated with the different $\boldsymbol{(g,h)}$-sectors are non-zero}. 
\\[1ex] 
The partition function~\eqref{noncompactPI} associated with the non-compact Minkowskian 
space-time directions is real, modular invariant and non-vanishing by itself. Then, the only way 
that the full partition function $\mathcal{Z}_\text{full}$ in eq.~\eqref{FacFullPI} 
vanishes, is that the internal partition function 
$\mathcal{Z}_\text{int.}$ is zero. Since we assume 
$\mathcal{Z}_\text{int.}\brkt{g}{h} \neq 0$ for some $g,h \in \mathbf{S}$, this option 
involves a non-trivial cancellation between the contributions from the various twisted sectors in 
eq.~\eqref{IntTwistedPI}.
\item[{\bf 3.}] {\bf The twisted internal partition function $\boldsymbol{\mathcal{Z}_\text{int.}\brkt{g}{h}}$ in each $\boldsymbol{(g,h)}$-sector vanishes identically by itself.}
}
Since both the first and the second option involve very delicate cancellations, which may be hard 
to realize in a model-independent way, we focus in this paper on the third option and demand that 
$\mathcal{Z}_\text{int.}\brkt{g}{h}$, given in eq.~\eqref{FacIntTwistedPI}, vanishes for all pairs 
$[g,h] = 0$. The ten-dimensional one-loop sector $(\Id,\Id)$ is modular invariant by itself. Hence, 
we need to require that it vanishes by itself. This leads us to restrict ourselves to one of the 
two supersymmetric heterotic string theories, i.e.\ the $\E{8}\times\E{8}$ or the 
$\Spin{32}/\Intr_2$ string, as the starting point of our investigation. Furthermore, if the 
geometrical point group $\mathbf{P}_\rep{v}$ and its realization $\mathbf{P}_\rep{s}$ on the 
spinors are not isomorphic, then $-\Id_8\in\mathbf{P}_\rep{s}$. Hence, we know immediately that 
there is some space group element, $g=(-\Id,\gm \, e)$, $\gm \in \Ratl^6$, that does not admit any 
invariant spinor. If $\gm=0$ then the ten-dimensional part of the string theory corresponds to one 
of the non-supersymmetric heterotic strings (i.e.\ either the tachyon-free $\SO{16}\times\SO{16}$ 
string~\cite{AlvarezGaume:1986jb} or one of tachyonic ones~\cite{Dixon:1986iz} depending on the 
choice of gauge embedding for this element $-\Id_8$). Moreover, if $\gm\neq 0$ then there is some 
Scherk-Schwarz supersymmetry breaking associated with some torus. Since in either case the internal 
partition function does not vanish in the sector $(g,\Id)$, we restrict ourselves to those cases 
where the abstract point group $\mathbf{P}$ and its realizations $\mathbf{P}_\rep{v}$ in geometry 
and $\mathbf{P}_\rep{s}$ in spinor space are all isomorphic. 
%

%
Now, since the twisted internal partition function~\eqref{FacIntTwistedPI}  for $(g,h)\neq(\Id,\Id)$ 
consists of a product of three parts, it vanishes whenever one of them vanishes. The $(g, h)$-twisted 
partition function $\mathcal{Z}_X\brkt{g}{h}$ associated with the internal orbifold geometry is typically 
non-zero: Only in very special points in the moduli space, where e.g.\ a free-fermionic equivalent 
description applies, it may vanish. Ignoring such special cases for symmetric orbifolds, the 
internal coordinate fields $X$ cannot make the partition function~\eqref{FacIntTwistedPI}  vanish.

\subsection{Vanishing Right-Moving Fermionic Partition Functions}

On the contrary, the right-moving fermionic partition function $\mathcal{Z}_\gps\brkt{g}{h}$ in eq.~\eqref{FacIntTwistedPI} can become zero under special circumstances: As we show in appendix~\ref{App:AppRiemannId} using a 
variant of the famous Riemann identities for products of four theta functions, this partition 
function vanishes if and only if both the constructing and projecting space group elements preserve 
at least one common spinor: 
\begin{align}\label{eq:VanishingCondition}
\mathcal{Z}_\gps\brkt{g}{h}(\tau)~=~0 \qquad \Leftrightarrow \qquad
\begin{array}{l}
\text{The space group elements $g, h \in \mathbf{S}$ admit }\\
\text{at least one common Killing spinor.}
\end{array} 
\end{align}
Indeed, for $g =(\gth, e\, m) \in \mathbf{S}$ the possible eigenvalues of $D_\rep{s}(\gth)$ 
are $\exp(\pm 2\pi i\, \widetilde{v}_g^a)$, $a=1,2,3,4$, where
\equ{ \label{KillingSpinorConditions}
\widetilde{v}_g ~=~ 
\frac 12 \pmtrx{
\phantom{-}v_g^1 + v_g^2 +  v_g^3 \\[1ex] 
-v_g^1 + v_g^2 +  v_g^3 \\[1ex] 
\phantom{-}v_g^1 - v_g^2 +  v_g^3 \\[1ex] 
\phantom{-}v_g^1 + v_g^2 -  v_g^3 
}~,
}
and similarly for $h=(\rho, e\, n) \in \mathbf{S}$ with $h\,g = g\,h$, see 
eq.~\eqref{eq:localTwistBasisB}. If the same components $\widetilde{v}_g^a$ and 
$\widetilde{v}_{h}^a$ vanish modulo integers, the right-moving fermionic partition function 
$\mathcal{Z}_\gps\brkt{g}{h}$ is zero, see eq.~\eqref{tildeLocalTwist}. Furthermore, in this 
case $D_\rep{s}(\gth)$ and $D_\rep{s}(\rho)$ have a common eigenvector with eigenvalue +1. In 
other words, there exists a common Killing spinor that is simultaneously invariant under 
$D_\rep{s}(\gth)$ and $D_\rep{s}(\rho)$. If a Killing spinor is invariant under 
at least one point group element $D_\rep{s}(\gth)$, we say that a Killing spinor exists 
\emph{locally} for $\gth \in \mathbf{P}$.
From this fact one realizes that the partition function of any supersymmetric orbifold always 
vanishes identically, because in this case there exists (at least) one globally defined constant 
spinor, meaning there is a spinor that is $D_\rep{s}(\gth)$-invariant for all $\gth \in \mathbf{P}$. 
In contrast to a Killing spinor being \emph{local}, we call such a Killing spinor \emph{global}\footnote{The phrase \emph{local} and \emph{global} Killing spinors refers to whether the Killing spinor 
is invariant under a single point group element or under the full point group, respectively. It is 
not related to local and global supersymmetry.}.

Similarly, we can arrange that the left-moving side $\mathcal{Z}_Y\brkt{g}{h}$ vanishes identically 
in eq.~\eqref{FacIntTwistedPI} by invoking some generalized Riemann identities involving more 
than four theta functions (like eq.~\eqref{VanishingPF} derived in appendix~\ref{ssc:GeneralizedIdentities}). 
Such identities rely on the fact that there are different contributions with the same mass that 
cancel among each other. As argued in appendix~\ref{app:LMfermiPartition}, since there is just a 
single left-moving vacuum state, it has to be projected out for the generalized Riemann identity to 
hold. Consequently, there are no massless gravitons or Cartan gauge fields in this setting, as they 
are obtained as oscillator excitations of this left-moving vacuum state.
This is in conflict with the physical requirement that at least the four-dimensional graviton and 
some (Cartan) gauge fields -- to be identified with, for example, some of the Standard Model gauge 
fields -- are part of the massless heterotic string spectrum. (If the graviton would not be part of 
the massless spectrum of string theory, why care about the cosmological constant in the first place.) 
For this reason, we reject the possibility to make $\mathcal{Z}_Y\brkt{g}{h}$ vanish identically 
and consequently assume that a systematic vanishing of $(g,h)$-twisted internal partition functions 
can only be guaranteed, if eq.~\eqref{eq:VanishingCondition} holds in all twisted sectors $(g,h)$. 

Obviously for toroidal orbifolds that preserve four-dimensional supersymmetry, the right-moving 
fermionic partition function $\mathcal{Z}_\psi\brkt{g}{h}$ vanishes identically for all commuting 
$(g,h)$-sectors. In principle, one could 
hope that there exist non-supersymmetric toroidal orbifold compactifications for which the 
right-moving partition function nevertheless vanishes in each $(g,h)$-twisted sector separately 
due to the existence of different supersymmetries in each $(g,h)$-sector. For this, it is 
sufficient that:
\begin{equation}\label{eq:OurGoal}
\hspace{-4ex}
\begin{minipage}{16cm}
\enums{
	\item[i.]  A Killing spinor exists \textit{locally} in every $(g,h)$-sector.
	In other words, each orbifold sector preserves by itself at least $\mathcal N=1$ space-time supersymmetry.
	\item[ii.] Not all $(g,h)$-sectors preserve the same Killing spinor(s),
        such that it is impossible to define any \textit{globally} invariant spinor.
	Hence, target-space supersymmetry is in total entirely broken.
}
\end{minipage}
\end{equation}
The main no-go result of this paper is that non-supersymmetric toroidal orbifolds with these 
properties do not exist. In fact, as we will show in the remainder of this paper, in each possible 
space group $\mathbf{S}$ associated with a non-supersymmetric toroidal orbifold, there is at least 
one element $g \in \mathbf{S}$ which by itself breaks all supersymmetries, thus violating point 1. 
In that case, the right-moving fermionic partition function $\mathcal{Z}_\psi\brkt{g}{h}$ receives 
a non-vanishing contribution precisely from such a $(g,\Id)$-sector, i.e.\ $h=\Id$ and $g\neq \Id$. 
Consequently, the only toroidal orbifold compactifications of the heterotic string for which 
$\mathcal{Z}_\psi\brkt{g}{h}$ vanishes identically for all commuting $(g,h)$-sectors actually 
preserve target-space supersymmetry.

\section{Nonexistence of Non-Supersymmetric Toroidal Orbifolds with Local Killing Spinors for Every Space Group Element}
\label{sec:ExplicitNoGo} 


\noindent 
In the previous section we saw that a necessary condition to have a large class of 
non-supersymmetric heterotic orbifold theories with vanishing cosmological constant is the 
following property: for each point group element separately some amount of supersymmetry is 
preserved but globally, i.e.\ for the full point group, no Killing spinor exists. 
In this section we will show that there are no such toroidal orbifolds.

\subsection{CARAT-Classification of Toroidal Orbifolds}


To show this statement we make use of the fact that all -- Abelian and non-Abelian -- point groups 
and all space groups relevant for toroidal orbifolds in six dimensions have been classified. 
Indeed, all space groups have been classified crystallographically in dimensions up to 
$D=6$~\cite{Opgenorth:1998ab,Plesken:2000ab}. In this classification, the geometrical space groups 
are sorted according to their geometrical point groups ($\Ratl$-classes), their lattices 
($\Z{}$-classes) and the possible roto-translations of the lattices (affine classes). 
It turns out that there are 7,103 $\Ratl$-classes, 85,308 $\Intr$-classes and a total of 28,927,915 
affine-classes; the latter label all geometrically inequivalent toroidal orbifolds in six dimensions. 
The collection of all this toroidal orbifold data is readily available electronically in the 
\textsc{carat} package~\cite{Opgenorth:1998ab}. Hence, we take this as our starting point in this 
section.


To understand the properties of possible local and global Killing spinors the $\Ratl$-classes are 
of particular interest. All inequivalent geometrical point groups that act 
crystallographically via eq.\ \eqref{eqn:chrystallographicaction} on some six-dimensional torus are 
labelled by their \textsc{carat} $\Ratl$-class index from 1 to 7,103. In each case, the generators 
of the geometrical point group are given in the lattice basis, denoted by 
$\widehat{\mathbf{P}}_\rep{v}$,  i.e.\ as a finite set of integral $6 \times 6$ matrices 
$\widehat{D}_\rep{v}(\gth) \in \widehat{\mathbf{P}}_\rep{v} \subset \text{GL}(6;\Intr)$. To find a 
representation of the corresponding group $\mathbf{P}_\rep{v} \subset \SO{6}$, we make use of the 
orthogonality property $D_\rep{v}(\gth)^{T}\, D_\rep{v}(\gth) = \Id_6$ to first identify the 
condition 
\begin{equation}
G ~=~ \widehat{D}_\rep{v}(\gth)^{\,T}\, G\, \widehat{D}_\rep{v}(\gth)~, 
\end{equation}
for all $\widehat{D}_\rep{v}(\gth) \in \widehat{\mathbf{P}}_\rep{v}$, on the torus 
metric~\eqref{TorusMetric} keeping in mind eq.~\eqref{eqn:chrystallographicaction}. A solution to 
the equation above is given by
\begin{equation}\label{eq:ansatz_G}
G ~=~ \sum_{\widehat{\gr} \,\in\, \widehat{\mathbf{P}}_\rep{v}} \; \widehat{\gr}^{\,T} \widehat{\gr}~.
\end{equation}
Using a Cholesky decomposition of $G$ we find a lattice vielbein $e$ and, consequently, 
the orthogonal matrices $D_\rep{v}(\gth) = e\, \widehat{D}_\rep{v}(\gth)\, e^{-1}$ 
using eq.~\eqref{eqn:chrystallographicaction} again.
But as mentioned in the previous section, the geometrical point group $\mathbf{P}_\rep{v}$ 
thus determined, does not fully specify the action on the spinors because of the double cover 
ambiguities. Each orthogonal matrix $D_\rep{v}(\gth)$ can be written in terms of the Lie algebra as 
in eq.~\eqref{Def:VectorRep} of appendix~\ref{App:Spin}. Then, using eq.~\eqref{Def:SpinorRep}
we explicitly obtain both possible representations $\pm D_\rep{s}(\gth)$ in spinor space. 
Hence, there may be many different abstract point groups $\mathbf{P}$ associated to the 
same geometrical point group $\mathbf{P}_\rep{v}$ (as provided by \textsc{carat}).

\subsection{Counting Invariant Spinors}
\label{sec:PointGroupActionSpinors}


To define the number of globally and locally preserved Killing spinors and to 
understand their distinction better, let us first discuss how to determine the number of Killing 
spinors preserved by some subgroup $\mathbf{G} \subset \mathbf{P}$. For example, this may be a 
$\Z{N_\gth}$ subgroup $\mathbf{G} = \langle \gth \rangle \subset \mathbf{P}$ generated by any 
$\gth \in \mathbf{P}$ or the whole point group, $\mathbf{G} = \mathbf{P}$. Using a four-dimensional 
Weyl representation, $D_\rep{4}$, defined in appendix~\ref{ChiralBasis}, each $\mathbf{G}$-invariant 
Weyl spinor $\Psi_\text{inv.}$ satisfies the condition
\begin{equation}
D_\rep{4}(\gth)\, \Psi_\text{inv.} ~=~ \Psi_\text{inv.}~, 
\end{equation}
for all $\gth \in \mathbf{G}$. Consequently, the $\mathbf{G}$-invariant spinor eigenspace can be found 
using the projection operator
\begin{equation}
\mathcal{P}^{\mathbf{G}} ~=~ \frac{1}{|\mathbf{G}|}\sum_{\gth' \,\in\, \mathbf{G}} D_\rep{4}(\gth')~,
\end{equation}
which is defined such that $D_\rep{4}(\gth)\,\mathcal{P}^{\mathbf{G}} = \mathcal{P}^{\mathbf{G}}$ 
 for all $\gth \in \mathbf{G}$. Then, the number of $\mathbf{G}$-invariant Killing spinors is counted using the trace of the projection operator
\begin{equation}\label{eq:SUSYpreservedbyG_part1}
\mathcal{N}^{\mathbf{G}} ~=~ \text{Tr}\big( \mathcal{P}^{\mathbf{G}}\big) ~=~ 
\frac{1}{|\mathbf{G}|} \sum_{\gth' \,\in\, \mathbf{G}} \text{Tr}\left(D_\rep{4}(\gth')\right)~. 
\end{equation}
In particular, we say that $\mathcal{N}^{\langle \gth \rangle}$ determines the number of 
{\em local} Killing spinors compatible with the point group element $\gth \in \mathbf{P}$, while 
$\mathcal{N} = \mathcal{N}^{\mathbf{P}}$ gives the number of {\em global} Killing spinors and hence the amount of target-space supersymmetry.

\subsection{Point Groups Admitting Local Killing Spinors Without Global Ones}\label{sec:LocalNotGlobal}

%
\begin{table}[t]
\begin{center}
\renewcommand{\arraystretch}{1.2}
\begin{tabular}{| r | p{13cm} |}
\hline
{\bf \# $\Ratl$-classes} & {\bf Restriction} \\
\hline\hline 
7,103~~ & All inequivalent geometrical point groups $\mathbf{P}_\rep{v} \subset \text{O}(6)$ \\
1,616~~ & Orientable geometrical point groups $\mathbf{P}_\rep{v} \subset \SO{6}$ \\
106~~  & No element from $\mathbf{P}_\rep{v}$ rotates in a two-dimensional plane only \\
63~~   & Each element $\gth \in \mathbf{P}_\rep{v}$ admits a choice with $\mathcal{N}^{\langle \gth\rangle} \geq 1$ local Killing spinors \\
60~~   & Geometrical point group compatible with $\mathcal{N} \geq 1$ global Killing spinors \\
\hline
\end{tabular}
\renewcommand{\arraystretch}{1}
\end{center} 
\caption{ \label{tab:QClassesSUSY} 
The number of $\Ratl$-classes compatible with various necessary requirements for local and global Killing spinors on six-dimensional toroidal orbifolds. 
}
\end{table}


With these definitions in place we can investigate which of the $7,\!103$ classified geometrical 
point groups are relevant for the present investigation. As can be inferred from 
table~\ref{tab:QClassesSUSY}, the vast majority of $\Ratl$-classes has no chance to admit a local 
Killing spinor for each element, which is a necessary condition for a vanishing partition function.
Indeed, most geometrical point groups contain elements which do not preserve the orientation, i.e.\ 
$\mathbf{P}_\rep{v} \subset \text{O}(6)$ but $\mathbf{P}_\rep{v} \not\subset \SO{6}$, hence do not 
allow for any definition of spinors in the first place. Out of the remaining point groups, 
again a large portion can be discarded because some elements rotate non-trivially in one complex 
plane only. In this case, the associated local twist vector $v_g$ defined in 
eq.~\eqref{eq:localTwistBasisA} can be brought to the form $v_g = (0,v_g^1,0,0)$. Consequently, 
none of the eigenvalues of the associated spinorial representation 
$D_\rep{s}(\gth)$ as given in eq.~\eqref{eq:localTwistBasisB} is one, i.e.\ none of the components 
of the vector defined in eq.~\eqref{KillingSpinorConditions} vanishes modulo integers. Thus, 
$D_\rep{s}(\gth)$ has no invariant Killing spinors. (This is in particular the case when 
$v_g^1=1$ corresponding to the twist $D_\rep{s}(\gth)=-\Id_8$ on the fermions.)

According to table~\ref{tab:QClassesSUSY} we are ultimately left with 106 geometrical point groups 
to be consider further. Out of these 106 candidates there are 63 $\Ratl$-classes with the following 
property: For every geometrical point group element $D_\rep{v}(\gth) \in \mathbf{P}_\rep{v}$ there 
is an embedding in $\text{Spin}(6)$, such that $D_\rep{s}(\gth)$ separately admits Killing spinors. 
This means that all elements of the associated $\mathbf{P}_\rep{s}$ can individually lie inside an 
$\SU{3}$ and thus all have special holonomy. To perform such a check for the non-trivial $\SU3$ 
holonomy we partly recycle the methods developed in~\cite{Fischer:2012qj,Fischer:2013qza} with the 
aim of searching for orbifolds that allow for a global Killing spinor. In contrast, here, we do not 
apply this \SU3-check to the whole point group but rather to each $\Z{N}$ subgroup generated by 
$D_\rep{v}(\gth) \in \textbf{P}_\rep{v}$. This check can be performed without constructing the 
possible spin embeddings, since the relevant conditions are formulated on the level of the 
geometrical point group $\mathbf{P}_\rep{v}$.


On the other hand, 60 out of 63 $\Ratl$-classes had been identified in \cite{Fischer:2012qj} to 
admit $\mathcal{N}\geq 1$ Killing spinors globally (for previous partial classifications 
see e.g.~\cite{Donagi:2004ht,Forste:2006wq,Donagi:2008xy}) and, hence, also locally. 
However, a priori it is not clear whether 
these 60 point groups also allow for another choice of the spin embedding $\mathbf{P}_\rep{s}$ 
such that different Killing spinors are preserved in the various sectors, thus breaking 
supersymmetry globally but keeping invariant Killing spinors locally. Consequently, we conclude 
that there could be at most 63 geometrical point groups with $\mathcal N=0$ target-space 
supersymmetry, yet admitting Killing spinors locally for each point group element.
Some elementary properties of the three additional $\Ratl$-classes are collected in 
table~\ref{tab:ParrQClasses}.
Even though each point group element can be embedded individually into spinor space so as to preserve 
at least one Killing spinor, not all the required choices can be made at the same time 
for all elements in $\mathbf{P}_\rep{s}$.

For each of the 63 geometrical point groups $\mathbf{P}_\rep{v}$ we went through all possible choices 
to embed it into spinor space. We observed that in all cases the resulting group $\mathbf{P}_\rep{s}$ 
either preserves target-space supersymmetry \emph{globally} or contains some elements that do not 
preserve any Killing spinor. Hence, there does not exist any non-supersymmetric orbifold for 
which all point group elements separately preserve some Killing spinors.

One way of proving this no-go result is to analyze this in detail for each of these point groups by 
explicitly constructing all their spin representations. We have performed this investigation but 
unfortunately it is not that illuminating as to how this result comes about. Nevertheless, we 
exemplify this analysis for the {$Q_8$ orbifolds in section~\ref{sec:Q8Example}. In the 
following section we prove our no-go result by using only representation theory of finite groups, 
i.e.\ without having to explicitly construct any spin representations of the point group.

\begin{table} 
\[
\renewcommand{\arraystretch}{1.4}
\begin{array}{|c|c|c|c|c|}
\hline
{\textsc{carat}}\textbf{-Index} &\textbf{Group} & \textbf{Generator Relations}                                & \textbf{Order} & \textbf{Local Twist Vectors} \\\hline\hline
 3375 & \text{Dic}_3 = \Intr_3 \rtimes \Intr_4  & \gth_1^4 = \gth_2^3 = \Id,~ \gth_2\,\gth_1\,\gth_2=\gth_1          & 12 & \big( \frac{1}{4}, \frac{1}{4},-\frac{1}{2}\big)~, \big( \frac{1}{3},-\frac{1}{3}, 0\big) \\ \hline
 5751 & Q_8                                     & \gth_1^4 = \Id,~\gth_1^2 = \gth_2^2,~\gth_1\,\gth_2\,\gth_1=\gth_2 &  8 & \big( \frac{1}{4}, \frac{1}{4},-\frac{1}{2}\big)~, \big( \frac{1}{4},-\frac{1}{4}, 0\big) \\ \hline 
 6737 & \text{SL}(2,3)                          & \gth_1^3 = \gth_2^4 = \Id,  (\gth_2\, \gth_1)^2 = \gth_1^2\,\gth_2 & 24 & \big( \frac{1}{3}, \frac{1}{3},-\frac{2}{3}\big)~, \big( \frac{1}{4},-\frac{1}{4}, 0\big) \\ \hline
\end{array}
\renewcommand{\arraystretch}{1}
\] 
\caption{ \label{tab:ParrQClasses}
Some defining properties of the three $\Ratl$-classes for which all elements have an $\SU{3}$ 
holonomy only separately. Since all these groups are non-Abelian, the two local twist vectors 
corresponding to the two generators are obtained from $D_\rep{v}(\gth_1)$ and $D_\rep{v}(\gth_2)$ 
in eq.~\eqref{eq:localTwistBasisA} diagonalized in two different bases.
}
\end{table}

\section{Finite Group Theoretical Non-Existence Proof}
\label{sec:FiniteGroupNoGo} 

Instead of proving the non-existence of non-supersymmetric toroidal orbifolds by explicitly 
constructing the relevant spinor representations such that each point group element admits a 
Killing spinor, we show this result here using finite group theory. (The relevant representation 
theory of finite groups is recalled in appendix~\ref{app:FiniteGroup} for completeness.) Moreover, 
we will do so in a way which is as much as possible independent of the results of 
section~\ref{sec:LocalNotGlobal}, i.e.\ we start again from all $\Ratl$-classes associated to 
six-dimensional toroidal orbifolds. 

\subsection{Relevant Four-Dimensional Representations of Finite Groups}


The 7,103 $\Ratl$-classes of six-dimensional orbifolds provided by \textsc{carat} correspond to 
only 1,594 different abstract point groups: For a given abstract group $\mathbf{P}$ there can exist 
several inequivalent realizations as integral $6\times 6$-matrices with different eigenvalues (e.g.\ 
the so-called $\Z{6}$-I and $\Z{6}$-II point groups both correspond to the abstract group $\Z{6}$). 
The spinorial point group $\mathbf{P}_\rep{s}$, i.e.\ the point group action on spinors, may have 
various non-isomorphic realizations because it involves the double cover. As stated in 
section~\ref{sec:DoubleCover}, since our focus is on orbifolds that always admit Killing spinors 
for all space group elements locally, we may restrict ourselves to those cases where the abstract 
group $\mathbf{P}$ and its realizations $\mathbf{P}_\rep{v}$ in geometry and $\mathbf{P}_\rep{s}$ 
in spinor space are all isomorphic. For this reason we may take the 1,594 different abstract finite 
groups $\mathbf{P}$, underlying the 7,103 $\Ratl$-classes, as our starting point.


Since $\text{Spin}(6)$ and $\SU4$ are isomorphic, the point group action in spinor space is encoded 
in a four-dimensional irreducible representation $\rep{4}$ of $\SU4$. (Indeed, a chiral projection 
of $D_\rep{s}(\gth)$ yields $D_\rep{4}(\gth)$ or its conjugate, see 
eq.~\eqref{ReductionChiralBasis}.) Hence, we are interested in four-dimensional, generally reducible 
representations\footnote{For notational simplicity, we do not make a distinction between 
four-dimensional representations of abstract finite groups and those of $\SU{4}$.} $\rep{4}$ of 
these 1,594 finite groups $\mathbf{P}$. 
 However, not any four-dimensional representation $\rep{4}$ of one of these finite groups 
 corresponds to a spinor representation that can be associated to a six-dimensional toroidal 
 orbifold: It might be a representation originating from $\U{4}$ instead of $\SU{4}$. 
Moreover, the resulting six-dimensional representation, 
\begin{align}\label{eq:6equal4times4}
\rep{6} ~=~ [\rep{4}]_2~, 
\end{align}
obtained by the two-times anti-symmetrized tensor product of $\rep{4}$ (see 
appendix~\ref{app:FiniteGroup}), might not correspond to one of the vector representations listed 
in the \textsc{carat} classification. To avoid these issues, two conditions need to be fulfilled:
\begin{equation}\label{eq:Conditions4Rep}
\hspace{-5ex}
\begin{minipage}{16cm}
\enums{
\item $\text{det}\left(D_\rep{4}(\gth)\right) = 1$ for all $\gth \in \mathbf{P}$~: \\[1ex] 
This determinant condition is fulfilled if the singlet representation obtained by a four-times 
anti-symmetrization of the $\rep{4}$, denoted by $[\rep{4}]_4$, is trivial, i.e.\ the corresponding 
characters are unity on all conjugacy classes: $\gch_{[\rep{4}]_4} = (1,\ldots, 1)$. This can be 
checked easily by using the character formula of four-times anti-symmetrized 
representations~\eqref{AntiSymCharacters}. 
\item The representation matrices associated to eq.~\eqref{eq:6equal4times4} are isomorphic to a 
$\Ratl$-class~: \\[1ex] 
Using the character formula of two-times anti-symmetrized representations~\eqref{AntiSymCharacters} 
the character $\gch_\rep{6}=\gch_{[\rep{4}]_2}$ can be computed. The resulting character values 
$\gch_\rep{6}$ should be equal to the character values $\gch_\rep{v}$ of the vector representations 
evaluated by tracing the corresponding $\widehat{D}_\rep{v}(\gth)\in\text{GL}(6;\Intr)$ of the 
\textsc{carat} $\Ratl$-classes (assuming that the same ordering of the conjugacy classes of 
$\mathbf{P}$ and $\mathbf{P}_\rep{v}$ has been used).
}
\end{minipage}
\end{equation}
Hence, by exploiting finite group characters both conditions can be checked without ever having to 
construct any representation matrices explicitly, neither $D_\rep{4}(\gth) \in \SU{4}$ nor 
$D_\rep{6}(\gth) \in \SO{6}$.

\subsection{Killing Spinors and Singlet Representations}

In section~\ref{sec:PointGroupActionSpinors} we counted the number of $\mathbf{G}$-invariant 
spinors $\mathcal{N}^\mathbf{G}$ by taking the trace of the corresponding projection 
operator~\eqref{eq:SUSYpreservedbyG_part1}. Using the definition of 
characters~\eqref{DefCharacter} and their inner products~\eqref{ChInner} this formula can be written as
\begin{equation}\label{eq:SUSYpreservedbyG_part2}
\mathcal{N}^{\mathbf{G}} 
~=~ \frac{1}{|\mathbf{G}|} \sum_{\gth' \,\in\, \mathbf{G}} \chi_\rep{4}(\gth') 
~=~ \langle \chi_\rep{4}, \chi_\rep{1} \rangle_{\mathbf{G}} 
~=~ n_\rep{1}^\mathbf{G}
~, 
\end{equation}
where we have used that the character of the trivial representation $\rep{1}$ is always unity. In 
the last step we inserted the character decomposition~\eqref{IrrepCharDecom} for the branching of 
$\rep{4}$ into irreducible representations of $\mathbf{G}$ and used the orthonormality of 
irreducible characters~\eqref{OrthoCharacters}. Hence, we see that the number of 
$\mathbf{G}$-invariant spinors is determined by the number $n_\rep{1}^\mathbf{G}$ of trivial 
singlet representations $\rep{1}$ in the decomposition of $\rep{4}$ into irreducible 
representations of $\mathbf{G}$.

\subsubsection*{Local Killing Spinors and Trivial \boldmath $\Intr_{N_\gth}$-Singlets}

Each element $\gth \in \mathbf{P}$ generates a $\Intr_{N_\gth}$ subgroup of $\mathbf{P}$. Hence, 
applying the results above for $\mathbf{G} =  \langle \gth \rangle$, we see that number of 
{\em local} Killing spinors preserved by $\gth$ is given by the number of trivial singlets 
\equ{\label{eq:SUSYpreservedbyTheta}
\mathcal{N}^{\langle \gth\rangle} ~=~ n_\rep{1}^{\langle \gth\rangle}~, 
}
in the decomposition of the $\rep{4}$ of $\mathbf{P} $ into irreducible representations of 
$\Intr_{N_\gth}$.

\subsubsection*{Global Killing Spinors and Trivial $\mathbf{P}$-Singlets} 

The number of {\em global} Killing spinors is determined by the number of singlets,  
\equ{ \label{eq:SUSYpreservedbyP}
\mathcal{N} ~=~ n_\rep{1}^{\mathbf{P}}~,
}
in $\rep{4}$ w.r.t.\ the whole group $\mathbf{P}$. Obviously, if the spinor representation 
$\rep{4}$ contains the trivial singlet $\rep{1}$ of $\mathbf{P}$, it branches to the trivial 
singlet for all subgroups, including the $\Intr_{N_\gth}$ subgroups generated by single elements 
$\gth\in\mathbf{P}$. Hence, if an orbifold admits global Killing spinors, it admits local Killing 
spinors as well.

\subsection{Nonexistence Proof by Checking All Orbifold Geometries}

In order to prove that there are no non-supersymmetric toroidal orbifolds that admit Killing 
spinors locally in all sectors, we implemented the following procedure: 

For each of the 1,594 different abstract groups $\mathbf{P}$ we considered all faithful (but in 
general reducible) four-dimensional representations $\rep{4}$ of $\mathbf{P}$. We required that 
they do not contain a trivial singlet representation of $\mathbf{P}$ so as to avoid having global 
Killing spinors by eq.~\eqref{eq:SUSYpreservedbyP}. 
Furthermore, if $\mathbf{P}$ is non-Abelian, we excluded the case where $\rep{4}$ 
decomposes into four one-dimensional representations by restricting to faithful 
representations: Otherwise, the representation $\rep{4}$ would only generate an Abelian subgroup 
of $\mathbf{P}$, which by itself is in the list of 1,594 different abstract groups. Hence, this 
case is already accounted for. Furthermore, cyclic $\Z{M}$ groups were also disregarded, since 
for cyclic groups a local Killing spinor for its generator immediately results in a global Killing 
spinor and the resulting model is supersymmetric. Next, we selected only those representations 
$\rep{4}$ satisfying the two conditions mentioned in~\eqref{eq:Conditions4Rep} to ensure that the 
matrix representations $D_\rep{4}$ can act on spinors and the resulting six-dimensional 
representations $\rep{6} =[\rep{4}]_2$ act crystallographically on six-dimensional tori. 
Finally, we constructed all $\Z{N} \subset \mathbf{P}$ subgroups. We showed that for each remaining 
representation $\rep{4}$, there is at least one cyclic subgroup, for which $\rep{4}$ does not 
contain the trivial $\Z{N}$-singlet representation. Consequently, for all non-supersymmetric 
six-dimensional toroidal orbifolds there is always a sector without any local Killing spinor.

\subsection{A Finite Group Conjecture}


In section~\ref{sec:ExplicitNoGo} and in the present section we have shown, both using explicit 
constructions and by using abstract group theory, respectively, that there does not exist any 
six-dimensional toroidal orbifold that admits Killing spinors for all point group elements but none 
globally. While implementing the group theoretical methods in the computer-aided checks mentioned 
above, it turns out that condition 2.\ in~\eqref{eq:Conditions4Rep} is, in fact, obsolete: Even if 
we do not require that the abstract group has a crystallographic action associated with 
some six-dimensional toroidal orbifold, still we could not find any groups that admitted Killing 
spinors for all group elements locally but none globally. Since the number of Killing spinors was 
counted by the number of trivial singlet representations, we arrive at the following conjecture: 


\begin{conjecture} \label{Conjecture}
There does not exist any finite group $\mathbf{H}$ that has a four-dimensional representation 
$D_\rep{4}$ with the following three properties:
\begin{enumerate}
\renewcommand{\theenumi}{\roman{enumi}}
\item $D_\rep{4}$ has a trivial determinant, i.e.\ $\det\left(D_\rep{4}(\gth)\right) = 1$ for all $\gth \in \mathbf{H}$~, 
\item $D_\rep{4}$ does not contain the trivial singlet representation of $\mathbf{H}$~, 
\item but the branchings of $D_\rep{4}$ to all $\Z{N} \subset \mathbf{H}$ subgroups always contain 
the trivial $\Z{N}$-singlet representation.  
\end{enumerate} 
\end{conjecture} 
%
%
We have checked this conjecture for the following two lengthy lists of finite groups: 
\begin{itemize}
\item all 1,594 different finite groups which originate from the 7,103 $\Ratl$-classes of \textsc{carat};
\item all finite groups of order up to 500 from the SmallGroups Library of \textsc{gap}~\cite{GAP4}, 
which amounts to $\mathcal{O}(100,000)$ finite groups,
\end{itemize}
where we have again excluded orbifolds with cyclic point groups, since for them local Killing 
spinors always imply global supersymmetry.
Even though the 1,594 finite groups are associated to the \textsc{carat} $\Ratl$-classes, we have 
not implied any connection to toroidal orbifolds while checking this conjecture for this set of 
finite groups. The orders of these groups range from order one for the trivial group to order 
103,680 for $\Z{2} \times (\text{O}(5,3) \rtimes \Z{2})$ with \textsc{carat}-index 2804. Moreover, 
443 groups of them are of order 501 or higher, hence are not captured by the second list.

For each group $\mathbf{H}$ from these lists we constructed every (in general reducible) 
four-dimensional representation $\rep{4}$ of $\mathbf{H}$ that does not contain a trivial singlet 
of $\mathbf{H}$ with $D_\rep{4}(\gr) \in \SU{4}$ for every element $\gr \in \mathbf{H}$. Then, for 
every $\Z{N}$ subgroup of $\mathbf{H}$ we checked the branching of the representation $\rep{4}$ 
into irreducible representations of $\Z{N}$. In each case, we found at least one $\Z{N}$ subgroup 
of $\mathbf{H}$ where the $\rep{4}$ of $\mathbf{H}$ does not contain any trivial singlet of 
$\Z{N}$. We take this as strong evidence for the conjecture above.


However, our conjecture crucially depends on the condition that we need a four-dimensional 
representation: If we consider five- instead of four-dimensional representations in 
Conjecture~\ref{Conjecture}, then there are finite groups that fulfill the corresponding conditions 
of the conjecture. An explicit example is provided in section~\ref{sec:ViolatingConjecture} based 
on the finite group $Q_8$. Also when the condition {\em i.} on the determinant is relaxed, we can 
construct another example of a finite group for which the other conditions of the conjecture can be 
satisfied, see appendix~\ref{app:T7Example}.

\subsection{Possible Extensions Beyond Symmetric Toroidal Orbifolds}
\label{sec:Extensions}

In this paper we obtained a non-existence result for symmetric toroidal orbifolds only. 
However, in light of the conjecture formulated above one may wonder whether it can be extended to 
more general settings.


The conjecture formulated above suggests that our nonexistence result extends to asymmetric 
heterotic toroidal orbifolds as well. The set of possible point groups that act on the right-movers 
becomes much larger for asymmetric orbifolds than for the symmetric case~\cite{GrootNibbelink:2017usl}. 
(For example, the largest Abelian point group for two-dimensional tori is $\Intr_6$ for symmetric 
constructions, while it can be $\Intr_{12}$ for asymmetric cases~\cite{Harvey:1987da,Dabholkar:1998kv}.) 
However, our conjecture suggests that also for asymmetric heterotic orbifolds the nonexistence 
outcome will stand, since we have confirmed it for all $\mathcal{O}(100,000)$ finite groups of 
order smaller or equal 500. Hence, non-supersymmetric asymmetric orbifolds with vanishing partition 
functions in all one-loop twisted sectors individually are also excluded for all point groups up to 
that order.


Moreover, one could consider a Calabi-Yau threefold, which admits some finite group of automorphisms
at some specific point in its moduli space. Now, it might be possible that it acts on spinors 
in such a way that it admits one or more Killing spinors that are not Killing spinors preserved by 
the Calabi-Yau itself. Or one could think about a variant of this, namely a $\text{K3}\times T^2$, 
which is orbifolded by some finite group, such that the various generators of this group 
preserve incompatible $\cN=1$ supersymmetries. Could such constructions lead to a heterotic model 
where the partition functions always vanish or would an extension of our no-go result apply here as 
well? 

The following argument suggests that our negative result even extends to such cases: Suppose 
that the Calabi-Yau (or the K3) admits deformations to some six- (or four-)dimensional 
toroidal orbifold. These deformations should not be obstructed by modding out the finite group 
of automorphisms. Then, modding out the automorphisms just leads to some (other) 
six-dimensional orbifold. But given that our non-existence proof included all non-supersymmetric 
toroidal orbifolds, it applies to these cases as well.

\section[{Examples with Q8 Point Group}]{Examples with \boldmath $Q_8$ Point Group}
\label{sec:Q8Example}

The purpose of this section is to illustrate various aspects of the abstract concepts, introduced in 
the previous sections, using toroidal orbifolds involving the non-Abelian group $Q_8$. To this end, we 
begin by giving some basic facts about the group $Q_8$ and its representations in 
section~\ref{sec:Q8Basics}. Then, in section~\ref{sec:SpinInterpretation4Q8} we illustrate the 
conditions discussed in section~\ref{sec:FiniteGroupNoGo} to allow for a spinorial interpretation 
of four-dimensional representations of $Q_8$. Since only toroidal $Q_8$ orbifolds with 
\textsc{carat}-index 5750 can fulfill all conditions simultaneously, we use this 
$\Ratl$-class to illustrate that there are four inequivalent embeddings of the same geometrical 
$Q_8$ action into spinor-space in section~\ref{sec:DoubleCoverAmbiguitiesQ8}: One leads to a 
supersymmetric construction, while the other three do not admit any Killing spinors globally. In 
line of our general no-go result, we find that none of these three non-supersymmetric orbifolds can 
give a systematic (i.e.\ model-independent) solution to the cosmological constant problem, as there 
are $Q_8$ group elements which remove all Killing spinors locally. In section~\ref{sec:NonisomorphicQ8} we 
give two examples involving $Q_8$ in which the geometrical point groups $\mathbf{P}_\rep{v}$ and their 
spinorial realizations $\mathbf{P}_\rep{s}$ are not isomorphic. The first example is an Abelian 
$\Z{2}\times\Z{2}$ orbifold with a non-Abelian $Q_8$ action on spinors. The second example involves 
one of the promising orbifold geometries identified in table~\ref{tab:QClassesSUSY} with 
\textsc{carat}-index 5751. Again in line with our general findings, in both examples we find that 
$-\Id_8 \in \mathbf{P}_\rep{s}$, which locally remove all Killing spinors.
In the final section, we illustrate the importance of our assumptions for conjecture~\ref{Conjecture} 
by considering five- instead of four-dimensional representations of $Q_8$.

\subsection[Group Q8 and Its Representations]{Group \boldmath$Q_8$ and Its Representations}
\label{sec:Q8Basics}


The quaternion group $Q_8 = \langle \gth_{1},\gth_{2}\rangle$ (with \textsc{gap}-ID [8, 4]) can be 
generated by two generators $\gth_1$ and $\gth_2$, which fulfill the following three defining 
conditions
\equ{ \label{eq:DefRelationsQ8} 
\gth_{1}^{4} ~=~ \Id,
\qquad 
\gth_{2}^{2} ~=~ \gth_{1}^{2},
\qquad 
\gth_{1}\,\gth_{2}\,\gth_1 =\gth_{2}~. 
}
Consequently, it has eight elements in total, which can be divided into five conjugacy classes:
\begin{equation}\label{eq:Q8cc}
[\Id] ~=~ \{\Id\}~,\quad  
[\gth_1] ~=~ \{\gth_1, \gth_1^3\}~,\quad 
[\gth_2] ~=~ \{\gth_2, \gth_2^3\}~,\quad 
[\gth_3] ~=~ \{\gth_3, \gth_3^3\}~,\quad 
[\gth_1^2] ~=~ \{\gth_1^2\}~.
\end{equation}
where we used $\gth_3 = \gth_1\gth_2$ as convenient short-hand notation.


Since the number of conjugacy classes equals the number of irreducible representations, the group 
$Q_8$ has five irreducible representations: The trivial one-dimensional representation is denoted 
by $\rep{1}_{++}$. Furthermore, there are three non-trivial one-dimensional representations, 
$\rep{1}_{+-}$, $\rep{1}_{-+}$, $\rep{1}_{--}$ and a single faithful two-dimensional 
representation $\rep{2}$. Explicitly, the matrix representations of the generators $\gth_a$ read
\equ{ \label{MatrixIrrepsQ8} 
D_{\rep{1}_{\textsc{ab}}}(\gth_1) ~=~ \textsc{a}\,1~, 
\quad 
D_{\rep{1}_{\textsc{ab}}}(\gth_2) ~=~ \textsc{b}\,1~, 
\quad 
D_{\rep{1}_{\textsc{ab}}}(\gth_3) ~=~ \textsc{a}\!\cdot\!\textsc{b}\,1
\quad\text{and}\quad
D_\rep{2}(\gth_a) ~=~ -\I\, \gs_a~,  
}
for $\textsc{a},\textsc{b}=\pm$ and in terms of the Pauli matrices $\gs_a$. The 
$2\times 2$-matrices, $D_\rep{2}(\gth_a)$, $a=1,2$, generate the $Q_8$ group as a subgroup of 
$\SU{2}$. To each irreducible representation $\rep{r}$ an irreducible character $\gch_{\rep{r}}$ is 
associated. Evaluating these irreducible characters $\gch_{\rep{r}}$ on the conjugacy classes 
$[\gth]$ according to eq.\ \eqref{ChValues} leads to the character table $T$ of $Q_8$ given in 
table~\ref{characterTableQ8}.

%
\begin{table}[t]
\[
\renewcommand{\arraystretch}{1.2}
\begin{array}{|r||r|}
\hline
T= \gch_{\rep{r}}([\gth])       &
[\gth] = \mtrx{ [\Id] & [\gth_1]~ & [\gth_2]~  & [\gth_3]~ 
                    & [\gth_1^2]~ }  
\\ \hline\hline    
\arry{r}{
\rep{r} = \rep{1}_{++}  \\ \rep{1}_{+-} \\ \rep{1}_{-+}  \\ \rep{1}_{--} \\  \rep{2}_{\phantom{--}} 
} \phantom{\, \, \,}
& 
\pmtrx{ 
 \phantom{-}1~ &  \phantom{-}1~ &  \phantom{-}1~ &  \phantom{-}1~ &  \phantom{-}1~ \\
 \phantom{-}1~ &  \phantom{-}1~ & -1~ &  -1~ & \phantom{-}1~ \\
 \phantom{-}1~ & -1~ &  \phantom{-}1~ &  -1~ & \phantom{-}1~ \\
 \phantom{-}1~ & -1~ & -1~ &  \phantom{-}1~ &  \phantom{-}1~ \\
 \phantom{-}2~ &  \phantom{-}0~ &  \phantom{-}0~ &  \phantom{-}0~ &  -2~ 
} \phantom{\,}
\\ \hline 
\end{array}
\]
\caption{ \label{characterTableQ8} 
Character table $T$ of the quaternion group $Q_{8}$. }
\end{table}
%


Any representation $\rep{R}$ of $Q_8$ can be decomposed in singlet and doublet representations as 
\begin{equation}\label{eq:Q8_spinorrepdecompositionR}
\rep{R} ~=~  
\bigoplus_{\textsc{a},\textsc{b}=\pm} (\rep{1}_{\textsc{ab}}\oplus)^{n_{(\textsc{ab})}} \ \oplus \ (\rep{2} \, \oplus)^{n_{(\rep{2})}}~.
\end{equation}
From this we compute the character of the representation $\rep{R}$ using table~\ref{characterTableQ8} 
and eq.~\eqref{IrrepCharDecom} such that we obtain
\equa{\label{eq:Q8generalcharactersofR}
\chi_{\rep{R}}{} &=~ 
\chi_{\rep{R}}\pmtrx{[\Id]\,,~ [\gth_1]\,,~ [\gth_2]\,,~ [\gth_3]\,,~ [\gth_1^2]  } 
\\ 
&=~ 
\left(
n_{(\rep{1}_*)}+2n_{(\rep{2})}\,,  
\sum\limits_{\textsc{a},\textsc{b}=\pm} \textsc{a}\, n_{(\textsc{ab})}\,, 
\sum\limits_{\textsc{a},\textsc{b}=\pm} \textsc{b}\, n_{(\textsc{ab})}\,, 
\sum\limits_{\textsc{a},\textsc{b}=\pm} \textsc{a}\!\cdot\!\textsc{b}\, n_{(\textsc{ab})}\,, 
\ n_{(\rep{1}_*)}-2n_{(\rep{2})}
\right)~,
\non 
}
where $n_{(\rep{1}_*)} = n_{(++)} + n_{(+-)} + n_{(-+)} + n_{(--)}$ is the total number of 
one-dimensional representations in the decomposition of $\rep{R}$. Hence, the total number of 
singlets and the number of doublet representations are given by
\equ{ \label{NumberSingletsDoubletsQ8}
n_{(\rep{1}_*)} ~=~ \frac{\chi_{\rep{R}}([\Id]) + \chi_{\rep{R}}([\gth_1^2])}{2}~, 
 \qquad 
n_{(\rep{2})} ~=~ \frac{\chi_{\rep{R}}([\Id]) - \chi_{\rep{R}}([\gth_1^2])}{4}~, 
}
i.e.\ they are determined by the character $\chi_{\rep{R}}$ evaluated on $[\Id]$ and 
$[\gth_1^2]$ only. On the other hand, the distribution over the various types of singlets is encoded in the 
character $\chi_{\rep{R}}$ evaluated on $[\gth_1], [\gth_2]$ and $[\gth_3]$.


Since there are only five irreducible representations of $Q_8$, the rules of tensor products of these irreducible representations are quite simple: 
\equ{ \label{TensorProdQ8irreps}
\rep{1}_{\textsc{ab}} \otimes \rep{1}_{\textsc{cd}} ~=~ \rep{1}_{\textsc{a}\cdot\textsc{c}\,\, \textsc{b}\cdot\textsc{d}}~, 
\qquad 
\rep{1}_{\textsc{ab}} \otimes \rep{2} ~=~ \rep{2}~, 
\qquad 
\rep{2}\otimes\rep{2} ~=~ \bigoplus_{\textsc{a},\textsc{b}=\pm} \rep{1}_{\textsc{ab}}~, 
\qquad 
[\rep{2}]_2 ~=~ \rep{1}_{++}~.
}
These properties can be readily verified using the character table~\ref{characterTableQ8} and the 
fact that characters specify the representations uniquely and turn tensor products and direct sums 
into ordinary products and sums of characters, see eq.\ \eqref{SumProdChar}. The second relation 
shows that any information about which singlet is involved, is completely washed out when tensoring 
a singlet with $\rep{2}$ because of the zeros of the character $\gch_\rep{2}$ for the conjugacy 
classes $[\gth_a]$. Moreover, since $\gch_{\rep{2}\otimes\rep{2}} = (4,0,0,0,4)$, it follows from 
eq.~\eqref{NumberSingletsDoubletsQ8} that there are no doublets in $\rep{2}\otimes\rep{2}$. Then, 
adding the character contributions of all singlets, shows that this tensor product can be written 
as the direct sum of all four irreducible singlet representations. Similarly, the final relation of 
eq.~\eqref{TensorProdQ8irreps} can be confirmed via eq.~\eqref{AntiSymCharacters} and 
corresponds to $\text{det}(D_\rep{2}(\gth_a))=+1$. 

%
\begin{table} 
\[
\renewcommand{\arraystretch}{1.2}
\begin{array}{|cc||c|ccc|c|}
\hline 
\multicolumn{2}{|c||}{\textbf{Subgroup}} &  \rep{1}_{++}  & \rep{1}_{+-} & \rep{1}_{-+} & \rep{1}_{--} & \rep{2} \\ 
\boldsymbol{$\Z{N}$} & \textbf{gen.} & \downarrow & \downarrow & \downarrow & \downarrow & \downarrow \\ \hline\hline  
\Z4 & \gth_1   & \rep{1}_0 & \rep{1}_0 & \rep{1}_2 & \rep{1}_2 & \rep{1}_1 \oplus \rep{1}_3 \\
\Z4 & \gth_2   & \rep{1}_0 & \rep{1}_2 & \rep{1}_0 & \rep{1}_2 & \rep{1}_1 \oplus \rep{1}_3 \\
\Z4 & \gth_3   & \rep{1}_0 & \rep{1}_2 & \rep{1}_2 & \rep{1}_0 & \rep{1}_1 \oplus \rep{1}_3 \\ \hline 
\Z2 & \gth_1^2 & \rep{1}_0 & \rep{1}_0 & \rep{1}_0 & \rep{1}_0 & \rep{1}_1 \oplus \rep{1}_1 \\ \hline 
\end{array}
\]
\caption{\label{tab:Q8branchings}
Branching of the irreducible representations of $Q_8$ into the irreducible representations 
$\rep{1}_q$ of the various $\Z{N}$ subgroups of $Q_8$, see eq.~\eqref{ZNirreprs} and eq.~\eqref{ZNbranching}. 
}
\end{table}
%


The quaternion group $Q_8$ has three maximal subgroups (generated by $\gth_1$, $\gth_2$ and $\gth_3$) 
which are all isomorphic to $\Z{4}$. In addition, $\gth_1^2 = \gth_2^2 = \gth_3^2$ 
generate a $\Z{2}$ subgroup. The branching of the irreducible representations of $Q_8$ into 
irreducible representations of those $\Z{4}$ and $\Z{2}$ subgroups is given in table~\ref{tab:Q8branchings}.

\subsection[Spinorial Interpretation of 4 of Q8]{Spinorial Interpretation of $\rep{4}$ of $\boldsymbol{Q_8}$} 
\label{sec:SpinInterpretation4Q8} 

In this paper we have given a number of criteria on four-dimensional representations of an
abstract group $\mathbf{P}$ to ensure that a heterotic orbifold can be obtained which admits 
Killing spinors locally. This section applies these conditions to the possible $\rep{4}$ of $Q_8$. 
To exemplify the consequences of each condition, we first consider each condition separately and 
then all of them combined.

%
\begin{table}[t]
\[
\renewcommand{\arraystretch}{1.1}
\begin{array}{|c||c|c|r|r|c|} 
\hline
\boldsymbol{$\Ratl$}\textbf{-class} & \multicolumn{2}{c|}{\textbf{Lattice Basis Generators}} & 
\multicolumn{2}{c|}{\textbf{Det}} & 
\textbf{Decomposition of }\rep{v} \\
\textsc{carat}\textbf{-ind.} &\widehat{\gr}_1=\widehat{D}_\rep{v}(\gth_1) & \widehat{\gr}_2=\widehat{D}_\rep{v}(\gth_2) 
& |\widehat{\gr}_1| & |\widehat{\gr}_2| 
& \textbf{and its Character} \\ \hline\hline
5750 &
\scalebox{.6}{$\left(\begin{array}{rrrrrr}
-1 & -1 & -1 &  1 & \phantom{-}0 & \phantom{-}0 \\[1ex]
 1 &  1 &  0 & -1 &  0 &  0 \\[1ex]
 1 &  1 &  0 &  0 &  0 &  0 \\[1ex]
 0 &  1 & -1 &  0 &  0 &  0 \\[1ex]
 0 &  0 &  0 &  0 &  1 &  0 \\[1ex]
 0 &  0 &  0 &  0 &  0 &   1 
\end{array}\right)$} & 
\scalebox{.6}{$\left(\begin{array}{rrrrrr}
 0 & -1 &  1 &  \phantom{-}0 &  \phantom{-}0 &  \phantom{-}0 \\[1ex]
 0 &  0 & -1 &  1 &  0 &  0 \\[1ex]
-1 &  0 & -1 &  1 &  0 &  0 \\[1ex]
-1 & -1 & -1 &  1 &  0 &  0 \\[1ex]
 0 &  0 &  0 &  0 &  1 &  0 \\[1ex]
 0 &  0 &  0 &  0 &  0 &  1 
\end{array}\right)$} & 1~ & 1~ &
\arry{l}{\rep{1}_{++} \oplus \rep{1}_{++} \oplus \rep{2} \oplus \rep{2} 
\\[2ex] \gch_\rep{v} = (6,2,2,2,-2)}
\\ \hline
5751 &
\scalebox{.6}{$\left(\begin{array}{rrrrrr}
 \phantom{-}0 &  \phantom{-}1 & -1 &  0 &  0 &  0 \\[1ex]
 0 &  0 &  1 & -1 &  0 &  0 \\[1ex]
 1 &  0 &  1 & -1 &  0 &  0 \\[1ex]
 1 &  1 &  1 & -1 &  0 &  0 \\[1ex]
 0 &  0 &  0 &  0 & -1 &  0 \\[1ex]
 0 &  0 &  0 &  0 &  0 & -1 
\end{array}\right)$} & 
\scalebox{.6}{$\left(\begin{array}{rrrrrr}
-1 & -1 & -1 &  1 &  \phantom{-}0 & \phantom{-}0 \\[1ex]
 1 &  1 &  0 & -1 &  0 &  0 \\[1ex]
 1 &  1 &  0 &  0 &  0 &  0 \\[1ex]
 0 &  1 & -1 &  0 &  0 &  0 \\[1ex]
 0 &  0 &  0 &  0 &  1 &  0 \\[1ex]
 0 &  0 &  0 &  0 &  0 &  1 
\end{array}\right)$} &  1~ & 1~ & 
\arry{l}{\rep{1}_{-+} \oplus \rep{1}_{-+} \oplus \rep{2} \oplus \rep{2} 
\\[2ex] \gch_\rep{v}=(6,-2,2-2,-2) } \\ \hline
6100 &
\scalebox{.6}{$\left(\begin{array}{rrrrrr}
 0 &  0 &  1 &  1 &  \phantom{-}0 &  \phantom{-}0 \\[1ex]
 1 &  0 &  0 &  1 &  0 &  0 \\[1ex]
-1 &  1 &  1 &  1 &  0 &  0 \\[1ex]
 0 & -1 & -1 & -1 &  0 &  0 \\[1ex]
 0 &  0 &  0 &  0 &  1 &  0 \\[1ex]
 0 &  0 &  0 &  0 &  0 &  1 
\end{array}\right)$} &
\scalebox{.6}{$\left(\begin{array}{rrrrrr}
 1 & -1 & -1 &  0 &  \phantom{-}0 &  \phantom{-}0 \\[1ex]
 1 & -1 & -1 & -1 &  0 &  0 \\[1ex]
 1 &  0 &  0 &  1 &  0 &  0 \\[1ex]
-1 &  1 &  0 &  0 &  0 &  0 \\[1ex]
 0 &  0 &  0 &  0 &  1 &  0 \\[1ex]
 0 &  0 &  0 &  0 &  0 & -1 
\end{array}\right)$} &  1~ & -1~ & 
\arry{l}{ \rep{1}_{++} \oplus \rep{1}_{+-} \oplus \rep{2} \oplus \rep{2} 
\\[2ex] \gch_\rep{v}=(6,2,0,0,-2) }  \\ \hline
6101 &
\scalebox{.6}{$\left(\begin{array}{rrrrrr}
 1 & -1 & -1 &  0 &  \phantom{-}0 &  0 \\[1ex]
 1 & -1 & -1 & -1 &  0 &  0 \\[1ex]
 1 &  0 &  0 &  1 &  0 &  0 \\[1ex]
-1 &  1 &  0 &  0 &  0 &  0 \\[1ex]
 0 &  0 &  0 &  0 &  1 &  0 \\[1ex]
 0 &  0 &  0 &  0 &  0 & -1 
\end{array}\right)$} &
\scalebox{.6}{$\left(\begin{array}{rrrrrr}
 0 &  0 &  1 &  1 &  0 &  0 \\[1ex]
 1 &  0 &  0 &  1 &  0 &  0 \\[1ex]
-1 &  1 &  1 &  1 &  0 &  0 \\[1ex]
 0 & -1 & -1 & -1 &  0 &  0 \\[1ex]
 0 &  0 &  0 &  0 & -1 &  0 \\[1ex]
 0 &  0 &  0 &  0 &  0 & -1 
\end{array}\right)$} & -1~ & 1~ &
\arry{l}{ \rep{1}_{+-} \oplus \rep{1}_{--} \oplus \rep{2} \oplus \rep{2} 
\\[2ex] \gch_\rep{v}=(6,0,-2,0,-2) }  \\ \hline
\end{array}
\renewcommand{\arraystretch}{1}
\]
\caption{ \label{tab:Q8FromCARAT}
This table lists the four $\Ratl$-classes corresponding the quaternion group $Q_8$. The first column gives their 
\textsc{carat}-index. The next columns provide their two generators in the lattice basis and their determinants. 
The decomposition~\eqref{eq:Q8_spinorrepdecompositionR} of the six-dimensional vector representations $\rep{v}$ and their characters $\gch_\rep{v}$ are given in the final column. 
}
\end{table}

\subsubsection*{Orientable Toroidal Orbifold}


First of all, there exist only four $\Ratl$-classes based on $Q_8$. Their \textsc{carat}-indices are 
5750, 5751, 6100 and 6101. Details are listed in table~\ref{tab:Q8FromCARAT}. The $\Ratl$-classes with 
\textsc{carat}-indices 6100 and 6101 have one generator with determinant $-1$. Consequently, the 
corresponding toroidal orbifold geometries do not admit spinors, only the other two $\Ratl$-classes 
with \textsc{carat}-indices 5750 and 5751 do.

\subsubsection*{Faithful Four-Dimensional Representation}

Any four-dimensional representation of $Q_8$ is reducible. Its 
decomposition~\eqref{eq:Q8_spinorrepdecompositionR} into singlets and doublets can be realized in 
three ways: 
\equ{ \label{Realizations4} 
n_{(\rep{1}_*)}=4~;~n_{(\rep{2})}=0~, 
\qquad 
n_{(\rep{1}_*)}=2~;~n_{(\rep{2})}=1~, 
\qquad 
n_{(\rep{1}_*)}=0~;~n_{(\rep{2})}=2~,  
}
ignoring the distinction between the different singlet representations for a moment. Since only 
$\rep{2}$ is a faithful irreducible representation of $Q_8$, the first option ($n_{(\rep{2})}=0$) 
is irrelevant.

\subsubsection*{Four-Dimensional Representation With Unit Determinant: $\boldsymbol{D_{\rep{4}}(Q_8)\subset \SU{4}}$}

In order that the four-dimensional representation can be interpreted as a spinor representation, it 
is necessary that $D_{\rep{4}}(\gth) \in \SU{4}$ for all $\gth\in Q_8$. As discussed in point 1.\ 
of~\eqref{eq:Conditions4Rep} this can be tested by checking that the character of $\gch_{[\rep{4}]_4}([\gth])=1$ 
for all five conjugacy classes~\eqref{eq:Q8cc}. For all possible realizations~\eqref{Realizations4} 
it turns out that $\gch_{[\rep{4}]_4}([\Id])=\gch_{[\rep{4}]_4}([\gth_1^2])=1$. As observed above, 
for $\rep{4}= \rep{2}\oplus\rep{2}$ we have $\gch_{\rep{4}}([\gth_a])=0$. Hence, 
it follows from eq.~\eqref{AntiSymCharacters}, that also the other three $\gch_{[\rep{4}]_4}([\gth_a]) =1$. 
Therefore, 
\equ{ \label{4equals2x2}
\rep{4} ~=~ \rep{2}\oplus\rep{2}
}
defines a four-dimensional representation which admits a spinor interpretation. 

For the other faithful realization of four-dimensional $Q_8$ representations, $\rep{1}_{\textsc{ab}}\oplus\rep{1}_{\textsc{cd}}\oplus\rep{2}$,  requiring that $\gch_{[\rep{4}]_4}([\gth_a]) =1$ for all $a=1,2,3$ leads to three equations
\equ{
\gch_\rep{4}([\gth_a])^4 + 8\, \gch_\rep{4}([\gth_a])^2 - 48 ~=~ 0~.
}
Since all characters of $Q_8$ are real, this equation is solved by 
$\gch_\rep{4}([\gth_a]) = +2$ or $-2$ for $a=1,2,3$. Then, the explicit expressions for 
$\gch_\rep{4}([\gth_a])$ as given in eq.~\eqref{eq:Q8generalcharactersofR} imply that both singlet 
representations have to be the same, i.e.
\equ{ \label{eq:Q85750rep4irreps}
\rep{4} ~=~ \rep{1}_{\textsc{ab}} \oplus \rep{1}_{\textsc{ab}} \oplus \rep{2}~, 
}
for $\textsc{a},\textsc{b}=\pm$. Hence, there exist four such realizations.

\subsubsection*{Resulting Six-Dimensional Representation $\boldsymbol{\rep{6}=[\rep{4}]_2}$ Defines a $\Ratl$-class}


Table~\ref{tab:Q8FromCARAT} lists the four inequivalent realizations of $Q_8$ as crystallographic point groups (i.e.\ at the level of $\Ratl$-classes), providing their lattice generators, their decompositions into irreducible $Q_8$ representations and their characters. 
For each six-dimensional representation $\rep{v}$ listed in this table we try to find a corresponding four-dimensional representation $\rep{4}$ of $Q_8$ that yields this $\rep{v}$ via the relation $\rep{6}=[\rep{4}]_2$ and the character relation~\eqref{AntiSymCharacters}. 
To do so, we first apply the general decomposition~\eqref{eq:Q8_spinorrepdecompositionR} to $\rep{4}$, determine the corresponding character values $\gch_\rep{4}$ given in eq.~\eqref{eq:Q8generalcharactersofR} and insert them in eq.~\eqref{AntiSymCharacters} for $[\rep{4}]_2$. The dimension of this representation is determined by the character evaluated in $\Id$ and obviously equals six: $\chi_{\rep{6}}([\Id]) = (4^2-4)/2 = 6$. 
Next, we observe that $\gch_{\rep{v}}([\gth_1^2]) = -2$ for all $Q_8$ $\Ratl$-classes listed in table~\ref{tab:Q8FromCARAT}. 
Using eq.~\eqref{eq:Q8generalcharactersofR} together with $\chi_{\rep{4}}([\gth_1^4]) = \chi_{\rep{4}}([\Id]) = 4$ we investigate how this can be obtained:  
\begin{equation} \label{eq:Q8eqn1}
\chi_{\rep{6}}([\gth_1^2]) ~=~ \sfrac12\, \gch_\rep{4}([\gth_1^2])^{2} - 2 
~\stackrel{!}{=}~ -2~,
\end{equation}
which implies $\chi_{\rep{4}}([\gth_1^2]) = 0$. Consequently, {we have for the other three 
conjugacy classes $[\gth_a]$, $a=1,2,3$, of $Q_8$: 
\equ{ \label{eqs:Q8Conditions}
\chi_{\rep{6}}([\gth_a]) ~=~  \sfrac 12\, \chi_{\rep{4}}([\gth_a])^2~,   
}
using $\gth_a^2 = \gth_1^2$. Note that the right-hand-side is non-negative. Hence, when the 
six-dimensional representation $\rep{v}$ has negative character values for $\gth_a$, this analysis 
implies that there is no four-dimensional representation from which it can be obtained. Looking at 
table~\ref{tab:Q8FromCARAT}, we thus infer that the vector representations defined by the 
$\Ratl$-classes 5751 and 6101 cannot be obtained from any four-dimensional representation of $Q_8$. 

\subsubsection*{Combined Consequences}

We have seen that a four-dimensional representation of $Q_8$ that admits a spinorial interpretation 
yields two options: Either it is a direct sum of two $\rep{2}$ or of one $\rep{2}$ and two identical 
singlets. For $\rep{4} = \rep{2}\oplus\rep{2}$ we get $\gch_\rep{4} = (4,0,0,0,-4)$. Hence, 
$\gch_\rep{4}([\gth_1^2]) = -4 \neq 0$ and this option cannot be realized in any $Q_8$ toroidal orbifold 
geometry. However, the second option~\eqref{eq:Q85750rep4irreps}
has $\gch_\rep{4} = (4,\textsc{a}\!\cdot\!2,\textsc{b}\,2,\textsc{a}\!\cdot\!\textsc{b}\!\cdot\!2,0)$, especially 
$\gch_\rep{4}([\gth_1^2])=0$. Using eq.~\eqref{eqs:Q8Conditions} this yields 
$\chi_{\rep{6}}([\gth_a]) = 2$ for $a=1,2,3$, which coincides with the characters of the 
$\Ratl$-class with \textsc{carat}-index 5750, see table~\ref{tab:Q8FromCARAT}. 
Consequently, when we combine all these conditions together and insist that $Q_8$ is also the 
geometrical point group of a toroidal orbifold (i.e.\ $\mathbf{P}_\rep{v}$ and $\mathbf{P}_\rep{s}$ 
are isomorphic), we find that only the $\Ratl$-class with \textsc{carat}-index 5750 fulfills all 
conditions and the spinorial embedding has to be $\rep{4} = \rep{1}_{\textsc{ab}} \oplus \rep{1}_{\textsc{ab}} \oplus \rep{2}$.

\subsection{Consequences of the Double Cover Ambiguities}
\label{sec:DoubleCoverAmbiguitiesQ8} 

The analysis above showed that only the $Q_8$ $\Ratl$-class with \textsc{carat}-index 5750 has 
four-dimensional spinorial representations that can be associated with a $Q_8$ toroidal 
orbifold. Moreover, the decomposition of these four-dimensional representations is given by 
eq.~\eqref{eq:Q85750rep4irreps}. Since $\textsc{a},\textsc{b}=\pm$, there are four such 
four-dimensional representations. All of them correspond to the same six-dimensional 
representation of $Q_8$. In other words, the geometrical action of $Q_8$ can be embedded 
in four inequivalent ways into spinor space.
The origin of these four inequivalent embeddings can be understood as double cover ambiguities. 
Since $Q_8$ has two generators, $\gth_1$ and $\gth_2$, there are $2^2=4$ different matrix 
realizations of these generators in spinor space. Next, we follow the methods outlined in 
section~\ref{sec:ExplicitNoGo} to see this explicitly.

\subsubsection*{Explicit Spinorial Embeddings} 

From \textsc{carat} we get the $Q_8$ generators of the geometrical point group 
$\widehat{\mathbf{P}}_\rep{v}$ represented as GL$(6,\Z{})$ matrices in an unspecified 
lattice basis $e$, see table~\ref{tab:Q8FromCARAT} for \textsc{carat}-index 5750. To obtain the 
vielbein $e$, we first use eq.~\eqref{eq:ansatz_G} to determine a compatible metric $G$ and then 
apply a Cholesky decomposition. We find
\begin{equation}
G ~=~
\scalebox{.75}{$\left(
\begin{array}{rrrrrr} 
 16 & 8 & 8 & -8 & 0 & 0 \\[1ex]
 8 & 16 & 0 & -8 & 0 & 0 \\[1ex]
 8 & 0 & 16 & -8 & 0 & 0 \\[1ex]
 -8 & -8 & -8 & 16 & 0 & 0 \\[1ex]
 0 & 0 & 0 & 0 & 8 & 0 \\[1ex]
 0 & 0 & 0 & 0 & 0 & 8 
\end{array}
\right)$}~,
\qquad
e ~=~ 2\,
\scalebox{.75}{$\left(
\begin{array}{rrrrrr}
 2 & 1 & 1 & -1 & 0 & 0 \\[1ex]
 0 & \sqrt{3} & -\frac{1}{\sqrt{3}} & -\frac{1}{\sqrt{3}} & 0 & 0 \\[1ex]
 0 & 0 & 2 \sqrt{\frac{2}{3}} & - \sqrt{\frac{2}{3}} & 0 & 0 \\[1ex]
 0 & 0 & 0 &  \sqrt{2} & 0 & 0 \\[1ex]
 0 & 0 & 0 & 0 &  \sqrt{2} & 0 \\[1ex]
 0 & 0 & 0 & 0 & 0 &  \sqrt{2} 
\end{array}
\right)$}~.
\end{equation}
By performing the conjugation~\eqref{eqn:chrystallographicaction} we obtain the 
$\SO{6}$-representation matrices, 
\begin{equation}
D_\rep{v}(\gth_1) ~=~
\scalebox{.75}{$\left(
\begin{array}{rrrrrr}
 0 & -\frac{1}{\sqrt{3}} & -\sqrt{\frac{2}{3}} & 0 & 0 & 0 \\[1ex]
 \frac{1}{\sqrt{3}} & 0 & 0 & -\sqrt{\frac{2}{3}} & 0 & 0 \\[1ex]
 \sqrt{\frac{2}{3}} & 0 & 0 & \frac{1}{\sqrt{3}} & 0 & 0 \\[1ex]
 0 & \sqrt{\frac{2}{3}} & -\frac{1}{\sqrt{3}} & 0 & 0 & 0 \\[1ex]
 0 & 0 & 0 & 0 & 1 & 0 \\[1ex]
 0 & 0 & 0 & 0 & 0 & 1 
\end{array}
\right)$}~,
\qquad
D_\rep{v}(\gth_2) ~=~
\scalebox{.75}{$\left(
\begin{array}{rrrrrr}
 0 & -\frac{1}{\sqrt{3}} & \frac{1}{\sqrt{6}} & \frac{1}{\sqrt{2}} & 0 & 0 \\[1ex]
 \frac{1}{\sqrt{3}} & 0 & -\frac{1}{\sqrt{2}} & \frac{1}{\sqrt{6}} & 0 & 0 \\[1ex]
 -\frac{1}{\sqrt{6}} & \frac{1}{\sqrt{2}} & 0 & \frac{1}{\sqrt{3}} & 0 & 0 \\[1ex]
 -\frac{1}{\sqrt{2}} & -\frac{1}{\sqrt{6}} & -\frac{1}{\sqrt{3}} & 0 & 0 & 0 \\[1ex]
 0 & 0 & 0 & 0 & 1 & 0 \\[1ex] 
 0 & 0 & 0 & 0 & 0 & 1 
\end{array}
\right)$}~, 
\end{equation}
in the Euclidean basis. From this we can compute the angles appearing in the 
parametrization~\eqref{Def:VectorRep} for both generators in the vector representation. Inserting 
them into eq.~\eqref{Def:SpinorRep} leads to the eight-dimensional spinor representation 
$D_{\rep{s}}$. At this point, it is convenient to use the projector $P^{(+)}$ introduced in 
eq.~\eqref{eq:weylprojector} to express the obtained spinor representation in the Weyl-basis,
as a four-dimensional matrix representation, namely
\begin{equation} \label{Explicit4D4for5750}
D_{\rep{4}_{\textsc{ab}}}(\gth_{1}) ~=~ \textsc{a}\, 
\scalebox{.75}{$\left(
\begin{array}{cccc}
 1 & 0 & 0 & 0 \\[1ex]
 0 & -\frac{\I}{\sqrt{3}} & \sqrt{\frac{2}{3}} & 0 \\[1ex]
 0 & -\sqrt{\frac{2}{3}} & \frac{\I}{\sqrt{3}} & 0 \\[1ex]
 0 & 0 & 0 & 1 
\end{array}
\right)$}~,
\qquad
D_{\rep{4}_{\textsc{ab}}}(\gth_{2}) ~=~ \textsc{b}\, 
\scalebox{.75}{$\left(
\begin{array}{cccc}
 1 & 0 & 0 & 0 \\[1.5ex] 
 0 & -\frac{\I}{\sqrt{3}} & -\frac{\I}{\sqrt{2}}\!-\!\frac{1}{\sqrt{6}} 
 & 0 \\[1.5ex] 
 0 & -\frac{\I}{\sqrt{2}}\!+\!\frac{1}{\sqrt{6}} & \frac{\I}{\sqrt{3}} & 0 \\[1.5ex] 
 0 & 0 & 0 & 1 
\end{array}
\right)$}~.
\end{equation}
The signs $\textsc{a},\textsc{b}=\pm$ parameterize the four double cover ambiguities. (The overall 
signs $\textsc{a}$ and $\textsc{b}$ in eq.~\eqref{Explicit4D4for5750} do not alter the 
two-dimensional representation $\rep{2}$ from the non-trivial $2\times 2$ matrix blocks in 
$D_{\rep{4}_{\textsc{ab}}}(\gth_{a})$, since the tensor product $\rep{1}_{\textsc{ab}}$ times 
$\rep{2}$ equals $\rep{2}$ again, see eq.~\eqref{TensorProdQ8irreps}.) This precisely corresponds 
to the four inequivalent decompositions~\eqref{eq:Q85750rep4irreps} of $\rep{4}$ of $Q_8$ that 
admitted a spinorial interpretation. Hence, the abstract representation theory and the explicit 
construction of the spinorial representations yield exactly the same result.

\subsubsection*{Number of Local and Global Killing Spinors}

%
\begin{table}[t]
\[
\renewcommand{\arraystretch}{1.2}
\begin{array}{|c||c|c|c|c|c||c|} 
\hline
\textbf{Decomposition of} & \multicolumn{5}{c||}{\textbf{Number of Killing Spinors Locally}} & \textbf{Globally} 
\\ 
\textbf{the Spinorial Repr.} & \boldsymbol{\mathcal{N}^{\langle\Id\rangle}} & \boldsymbol{\mathcal{N}^{\langle\gth_{1}\rangle}} & \boldsymbol{\mathcal{N}^{\langle\gth_{2}\rangle}} & \boldsymbol{\mathcal{N}^{\langle\gth_3\rangle}} & \boldsymbol{\mathcal{N}^{\langle\gth^2_{1}\rangle}} & \boldsymbol{\mathcal{N}^{Q_{8}}} \\
\hline \hline
 \rep{4}_{++} ~=~ 
 \rep{2} \oplus \rep{1}_{++} \oplus \rep{1}_{++}  & 4 & 2 & 2 & 2 & 2 & 2 \\ 
 \rep{4}_{+-} ~=~  
 \rep{2} \oplus \rep{1}_{+-} \oplus \rep{1}_{+-} & 4 & 2 & 0 & 0 & 2 & 0 \\ 
 \rep{4}_{-+} ~=~  
 \rep{2} \oplus \rep{1}_{-+} \oplus \rep{1}_{-+} & 4 & 0 & 2 & 0 & 2 & 0 \\ 
 \rep{4}_{--}  ~=~ 
 \rep{2} \oplus \rep{1}_{--} \oplus \rep{1}_{--} & 4 & 0 & 0 & 2 & 2 & 0 \\ 
 \hline
\end{array}
\]
\caption{ \label{tab:LocalSUSYForQ8}
This table lists the four inequivalent four-dimensional spinorial representations $\rep{4}$ and 
their decompositions for the toroidal $Q_8$ orbifold with \textsc{carat}-index 5750. For each of 
them we give the number of Killing spinors preserved locally, i.e.\ by a representative of 
each of the five conjugacy classes $[\Id], [\gth_1], [\gth_2], [\gth_3]$ and $[\gth_1^2]$, and 
globally, i.e.\ by the point group $Q_8$ as a whole. 
}
\end{table}

Now, we can either use the explicit spinorial representation matrices~\eqref{Explicit4D4for5750} or 
the decompositions~\eqref{eq:Q85750rep4irreps} of the four-dimensional representation $\rep{4}$ of 
$Q_8$ to determine the number of Killing spinors, \emph{locally} and \emph{globally}. To do so we 
use eq.~\eqref{eq:SUSYpreservedbyTheta} applied to any representative of each of the five conjugacy 
classes and eq.~\eqref{eq:SUSYpreservedbyP} for the whole $Q_8$ point group. The 
results for the four inequivalent choices in the double cover are listed in 
table~\ref{tab:LocalSUSYForQ8}.

As can be inferred from the last column of this table, the choice $\textsc{ab}=++$ yields 
$\mathcal{N}=2$ supersymmetry. (This is consistent with the results of ref.~\cite{Fischer:2012qj} 
where all six-dimensional toroidal orbifolds were identified that preserve $\cN\geq 1$ supersymmetry.)  
The other three cases, $\textsc{ab}=+-,-+$ and $--$, yield $\mathcal{N}=0$ in four dimensions, as 
these $\rep{4}$'s do not contain any trivial singlet representation $\rep{1}_{++}$ of $Q_8$. For 
these cases there always exist two conjugacy classes, each generating a $\Z{4}$ subgroup of $Q_8$, 
which do not admit any Killing spinors locally. This shows the impossibility to construct a 
toroidal $Q_8$ orbifold without global Killing spinors, but where each group 
element by itself admits invariant spinors. Consequently, any non-supersymmetric orbifold 
constructed out of the previously analyzed $\mathbf{P}_\rep{v}$ and $\mathbf{P}_\rep{s}$ will generically suffer from the cosmological constant problem.

\subsection[Examples of Nonisomorphic Geometrical and Spinorial Point Groups]{Examples of Nonisomorphic Geometrical and Spinorial Point Groups}
\label{sec:NonisomorphicQ8}

At various places we have excluded situations where the geometrical point group $\mathbf{P}_\rep{v}$ 
and its spinorial realization $\mathbf{P}_\rep{s}$ were not isomorphic, since this implies 
$-\Id_8\in \mathbf{P}_\rep{s}$ and thus breaks all supersymmetries (see the discussion around 
eq.~\eqref{ModificationsDefRelations}). In this section, we give two explicit examples to 
illustrate how this happens in detail.

\subsubsection*{\boldmath A $\Z{2}\times\Z{2}$ Orbifold With a $Q_8$ Action on Spinors}

Let us analyze the decomposition~\eqref{4equals2x2} of the four-dimensional representation $\rep{4}$ 
of $Q_8$ into two doublets. Using eq.~\eqref{MatrixIrrepsQ8} it can be given explicitly as
\begin{equation} \label{Z2xZ2Q8OrbiS}
D_{\rep{4}}(\gth_{1}) ~=~ 
\left(
\begin{array}{rrrr}
- \I \gs_{1} & 0  \\ 
 0        & -\I \gs_{1} 
\end{array}
\right)~,
\qquad
D_{\rep{4}}(\gth_2) ~=~ 
\left(
\begin{array}{rrrr}
 -\I \gs_{2} & 0 \\ 
  0      & - \I \gs_{2}     
\end{array}
\right)~.
\end{equation}
This representation admits a spinorial interpretation. However, it cannot be associated to any 
geometrical $Q_8$ orbifold, see the $\Ratl$-classes listed in table~\ref{tab:Q8FromCARAT} and the 
discussion at the end of section~\ref{sec:SpinInterpretation4Q8}. Indeed, this spinorial 
representation does not lead to a faithful six-dimensional representation $\rep{6} = [\rep{4}]_2$ 
of $Q_8$, as we show in the following: The anti-symmetrized tensor product $[\rep{4}]_2$ leads to a 
direct sum of $Q_8$ singlet representations, i.e.
\equ{ 
[\rep{2}\oplus\rep{2}']_{2} ~=~ [\rep{2}]_2 \oplus [\rep{2}']_2 \oplus (\rep{2}\otimes\rep{2}') ~=~
\rep{1}_{++} \oplus \rep{1}_{++} \oplus \rep{1}_{++} \oplus 
\rep{1}_{+-} \oplus \rep{1}_{-+} \oplus \rep{1}_{--}~,
}
using the tensor products~\eqref{TensorProdQ8irreps}. (To evaluate the tensor product we labeled 
the second two-dimensional representation as $\rep{2}'$.) Using eq.~\eqref{MatrixIrrepsQ8} an 
explicit matrix representation of $\rep{6} = [\rep{4}]_2$ is given by 
\begin{equation} \label{Z2xZ2Q8OrbiV}
D_\rep{v}(\gth_1) ~=~
\scalebox{.75}{$\left(
\begin{array}{rrrrrr}
 \phantom{-}1 & 0 & 0 & 0 & 0 & 0 \\[1ex]
 0 &  \phantom{-}1 & 0 & 0 & 0 & 0 \\[1ex]
 0 & 0 &  \phantom{-}1 & 0 & 0 & 0 \\[1ex]
 0 &  0 & 0 & \phantom{-}1 & 0 & 0 \\[1ex]
 0 & 0 & 0 & 0 & -1 & 0 \\[1ex]
 0 & 0 & 0 & 0 & 0 & -1 
\end{array}
\right)$}~,
\qquad
D_\rep{v}(\gth_2) ~=~
\scalebox{.75}{$\left(
\begin{array}{rrrrrr}
  \phantom{-}1 & 0 & 0 & 0 & 0 & 0 \\[1ex]
 0 &  \phantom{-}1 & 0 & 0 & 0 & 0 \\[1ex]
 0 & 0 &  \phantom{-}1 & 0 & 0 & 0 \\[1ex]
 0 &  0 & 0& -1 & 0 & 0 \\[1ex]
 0 & 0 & 0 & 0 &  \phantom{-}1 & 0 \\[1ex]
 0 & 0 & 0 & 0 & 0 & -1 
\end{array}
\right)$}~.
\end{equation}
Consequently, the geometrical point group is $\Z{2}\times\Z{2}$ (the corresponding $\Ratl$-class 
has \textsc{carat}-index 4618), while the spinorial point group is $Q_8$. 

This example clearly illustrates that the geometrical and spinorial point groups need not be 
isomorphic: The four-dimensional matrix representation~\eqref{Z2xZ2Q8OrbiS} satisfies the following 
algebraic relations 
\equ{ \label{Z2xZ2Q8DefS} 
D_\rep{4}(\gth_1)^2 ~=~ -\Id_4~, 
\quad 
D_\rep{4}(\gth_2)^2 ~=~ -\Id_4~, 
\quad 
D_\rep{4}(\gth_1)\, D_\rep{4}(\gth_2) ~=~ -D_\rep{4}(\gth_2)\, D_\rep{4}(\gth_1)~,
}
which are equivalent to the defining relations of $Q_8$ as given in eq.~\eqref{eq:DefRelationsQ8}.
On the other hand, the corresponding six-dimensional matrix representation~\eqref{Z2xZ2Q8OrbiV} 
satisfies
\equ{ \label{Z2xZ2Q8DefV} 
D_\rep{v}(\gth_1)^2 ~=~ \Id_6~, 
\quad 
D_\rep{v}(\gth_2)^2 ~=~ \Id_6~, 
\quad 
D_\rep{v}(\gth_1)\, D_\rep{v}(\gth_2) ~=~ D_\rep{v}(\gth_2)\, D_\rep{v}(\gth_1)~. 
}
In this case all defining relations of $D_\rep{4}$ and $D_\rep{v}$ differ by minus-signs. 
Hence, we have an Abelian $\Z{2}\times\Z{2}$ orbifold at the level of the geometry with a 
non-Abelian $Q_8$ action on spinors. Moreover, since $\mathbf{P}_\rep{v}$ and 
$\mathbf{P}_\rep{s}$ are not isomorphic, we see explicitly that all supersymmetries are broken by 
$-\Id_8 \in \mathbf{P}_\rep{s}$.

\subsubsection*{\boldmath A $Q_8$ Orbifold with a $\Z{4}\rtimes\Z{4}$ Action on Spinors}

The next example treats one of the three special $\Ratl$-classes identified in table~\ref{tab:ParrQClasses} 
of section~\ref{sec:ExplicitNoGo}: The $Q_8$ orbifold with \textsc{carat}-index 5751. In this example 
the local twist vectors associated to all elements of $Q_8$ can be chosen individually such that 
they admit Killing spinors locally. Indeed, the local twist vectors defined in 
eq.~\eqref{eq:localTwistBasisA} are given by 
\equ{ 
\renewcommand{\arraystretch}{1.2}
\begin{array}{|c||c|c|c|c|} \hline 
[\gth] & [\gth_1]                           & [\gth_2]                      & [\gth_3]                           & [\gth_1^2]                    \\ \hline\hline
     v & \big(\frac14,\frac14,-\frac12\big) & \big(\frac14,-\frac14 ,0\big) & \big(\frac14,\frac14,-\frac12\big) & \big(\frac12,-\frac12 ,0\big) \\ \hline 
\end{array} 
}
for the four non-trivial conjugacy classes of $Q_8$ using a different basis to diagonalize 
$D_\rep{v}(\gth)$ in each case. One sees that all twist vectors preserve some supersymmetries 
individually. However, eq.~\eqref{eqs:Q8Conditions} implies that all $\gch_\rep{6}([\gth_a])\geq 0$, 
while we see from table~\ref{tab:Q8FromCARAT} that some $\gch_\rep{6}([\gth])$ are negative. Thus, 
we conclude that there is no isomorphic spin-embedding of $Q_8$ for the six-dimensional 
representation $\rep{6}$ of $Q_8$. One can also see this explicitly as follows:

We can apply the methods of section~\ref{sec:ExplicitNoGo} to determine the corresponding spinor 
representation matrices $D_\rep{4}(\gth)$. In accordance with 
eq.~\eqref{ModificationsDefRelations} some of the defining relations of $Q_8$ are modified 
from eq.~\eqref{eq:DefRelationsQ8} to 
\equ{ \label{Q8spinRel} 
D_\rep{4}(\gth_1)^4 ~=~ \Id_4~, 
\qquad 
D_\rep{4}(\gth_2)^2 ~=~ - D_\rep{4}(\gth_1)^2~, 
\qquad 
D_\rep{4}(\gth_1)\, D_\rep{4}(\gth_2)\, D_\rep{4}(\gth_1) ~=~ -D_\rep{4}(\gth_2)~,
}
which is $\Z{4}\rtimes\Z{4}$ with 16 elements.

The relations~\eqref{Q8spinRel} show that $-\Id_8 \in \mathbf{P}_\rep{s}$. Hence, there 
are point group elements that explicitly break all supersymmetries and generically the cosmological 
constant does not vanish for the corresponding heterotic orbifold models.

\subsection{Conjecture~\ref{Conjecture} with Five-Dimensional Representations is Violated}
\label{sec:ViolatingConjecture} 

To emphasize the importance of the conditions of conjecture~\ref{Conjecture} we give an example, 
which fulfills all but one of the requirements of conjecture~\ref{Conjecture}: We take a 
five-dimensional representation of $Q_8$ instead and show that one can easily construct a 
counter-example in this extended case: The five-dimensional representation,  
\begin{equation} \label{5repQ8}
\rep5 ~=~ \rep1_{+-} \oplus \rep1_{-+} \oplus \rep 1_{--} \oplus \rep2~, 
\end{equation}
has $\det(D_\rep{5}(\gth)) = 1$ for all $\gth \in Q_8$, since the two generators can be represented 
by the following matrices 
\equ{
D_\rep{5}(\gth_1) ~=~
\scalebox{.75}{$\left(
\begin{array}{rrrrr}
 \phantom{-}1 & 0 & 0 & 0 & 0  \\[1ex]
 0 &  -1 & 0 & 0 & 0  \\[1ex]
 0 & 0 & -1 & 0 & 0  \\[1ex]
 0 &  0 & 0 & 0 & -\I  \\[1ex]
 0 & 0 & 0 & -\I & 0  
\end{array}
\right)$}~,
\qquad
D_\rep{5}(\gth_2) ~=~
\scalebox{.75}{$\left(
\begin{array}{rrrrr}
 -1 & 0 & 0 & 0 & 0  \\[1ex]
 0 &  \phantom{-}1 & 0 & 0 & 0  \\[1ex]
 0 & 0 & -1 & 0 & 0  \\[1ex]
 0 &  0 & 0& 0 & -1  \\[1ex]
 0 & 0 & 0 & \phantom{-}1 & 0 
\end{array}
\right)$}~,
}
using eq.~\eqref{MatrixIrrepsQ8}. This five-dimensional representation~\eqref{5repQ8} does not 
contain the trivial representation $\rep{1}_{++}$. However, for each non-trivial singlet 
$\rep{1}_{\textsc{ab}}$, $\textsc{ab}\neq ++$, there is a subgroup of $Q_8$ in which this singlet 
branches to the trivial singlet $\rep{1}_0$ of $\Z{4}$ (as can be seen from table~\ref{tab:Q8branchings}). 
Thus, this $\rep{5}$ of $Q_8$ contains a trivial $\Z{N}$ singlet when branched to each 
$\Z{N}$ subgroup of $Q_8$. Hence, we have shown that this five-dimensional representation of 
$Q_8$ does fulfill the three conditions of conjecture~\ref{Conjecture}, however extended to the 
case of five-dimensional representations instead of four-dimensional ones. Consequently, the assumption 
of four-dimensional representations is crucial to our conjecture~\ref{Conjecture}.

\section{Conclusion and Outlook}
\label{sec:conclusion}

\subsubsection*{Summary of the Non-Existence Result}


We have investigated conditions under which the one-loop cosmological constant vanishes for 
symmetric six-dimensional non-supersymmetric toroidal orbifolds in the context of the heterotic string. 
To ensure generic model-independent results, i.e.\ to obtain findings that do not rely 
on the details of the gauge embedding, we aimed to achieve this by requiring that the right-moving 
fermionic partition function vanishes in each (twisted) sector of the theory individually.  
The condition for this to happen is that all commuting pairs of constructing and projecting 
space group elements, $g$ and $h$, respectively, possess common Killing spinors but no Killing 
spinor survives globally. This situation could be seen as a non-supersymmetric setting, where 
each $(g,h)$-twisted sector reserves some supersymmetry. However, we find that there does not exist 
any six-dimensional toroidal orbifold which preserves some amount of supersymmetry for each space 
group element separately, but none globally. In other words, for any admissible space group there 
is at least one element $g$ that breaks supersymmetry completely and consequently the contribution 
from the $(g,\Id)$-twisted sector to the one-loop partition function does not vanish identically. 
Therefore, the main result of this paper is the following: only supersymmetric orbifold geometries 
in six dimensions can achieve that the one-loop right-moving fermionic partition function vanishes in all sectors individually.


We provide two independent methods to prove the non-existence of six-dimensional toroidal orbifolds 
for which all space group elements admit some Killing spinors, while no Killing spinors exist for 
the space group as a whole. In the first method we construct all possible spin embeddings for all 
point groups of all six-dimensional toroidal orbifolds. The second method uses the fact that the 
number of Killing spinors (preserved by a finite group) can be determined as the number of 
trivial singlet representations contained in the decomposition of four-dimensional representations 
of that finite group. The latter method has the main advantage that one does not have to construct 
the spin embedding explicitly and one can rely on elegant finite group theoretical results. 
However, both methods are based on the fact that six-dimensional toroidal orbifolds have 
been fully classified and hence we needed to simply go through the complete list of 
all relevant point groups.


Our second, finite group theoretical, proof of our non-existence result was extended to an interesting mathematical conjecture for finite groups, namely: There is no finite group possessing a four dimensional representation which has a trivial determinant and does not contain a trivial singlet, while all its branchings to cyclic subgroups have trivial singlet representations. 
This conjecture led us to speculate that our no-go result extends beyond toroidal orbifolds to e.g.\ orbifolds of Calabi-Yau manifolds, where different Killing spinors are preserved by the Calabi-Yau and the orbifolding.


The impossibility of constructing non-supersymmetric heterotic toroidal orbifold models with generically vanishing cosmological constant seems to be in sharp contrast to the perturbative situation on the type-II side especially if the duality between heterotic and type II string theories persists in the non-supersymmetric case as suggested in~\cite{Blum:1997gw}.  
Indeed, there exist non-supersymmetric string models on asymmetric orbifolds with one-loop vanishing cosmological constant on the type-II side~\cite{Kachru:1998hd,Shiu:1998he} (and more recent variations~\cite{Satoh:2015nlc}).
As discussed in ref.~\cite{Shiu:1998he} some of the different orbifold twists preserve either only some right- or some left-moving supersymmetry, so that supersymmetry is completely broken in the construction as a whole, yet ensuring that the corresponding twisted sectors in the partition function all vanish identically\footnote{The heterotic orbifold dual to the type-II construction has a non-vanishing 
cosmological constant which turns out to be non-perturbative in the coupling of the original 
type-II theory \cite{Harvey:1998rc}. Specifically, there appears a mismatch between the number of bosons and fermions 
in the heterotic dual, both in the untwisted and twisted sectors of the asymmetric orbifold.}.
In light of this left-right alternating pattern, one could suggest to preform a similar construction on the heterotic side. 
However, as argued in this paper, such an attempt is doomed to fail: It is possible to construct a left-moving partition function that vanishes by making use of (generalized) Riemann identities. But then the left-moving vacuum has been projected out and consequently it is impossible to obtain a massless graviton in the heterotic string spectrum.

\subsubsection*{Outlook}

The negative outcome of our classification induces a  significant constraint in the space of phenomenologically viable models which one could hope to obtain (in a straight-forward way) from string theory. As such, the results of this paper make any unanswered questions
much more interesting and pressing. Related searches could} therefore be extended in various directions:


The arguments at the end of section~\ref{sec:Extensions} strongly rely on the existence of an 
unobstructed orbifold limit. However, there exist many Calabi-Yau spaces that do not admit any 
toroidal orbifold interpretation. Yet, there is a second class of Calabi-Yau manifolds where 
our arguments should apply: Many Calabi-Yau manifolds (like complete intersection Calabi-Yaus or 
Calabi-Yaus based on weighted projected spaces) can be described by gauged linear sigma 
models (GLSMs) in appropriate limits~\cite{Witten:1993yc,Distler:1995mi}. 
In general, after gauge fixing the worldsheet gauge symmetry there might be some remaining 
$\Intr_M$-factors that encode the global structure of the resulting geometry. Now, suppose 
that the GLSM preserves additional discrete symmetries that admit other Killing spinors than the 
ones preserved by the $\Intr_M$-factors. Then, again our conjecture seems to suggest that there is 
a sector for which the right-moving fermionic partition function does not vanish and supersymmetry 
is explicitly broken. Since certain GLSMs possess limits in which they can be described by Gepner 
models, this suggests that our no-go result could even be extended to non-supersymmetric Gepner 
models~\cite{GatoRivera:2007yi,GatoRivera:2008zn}.


Both the toroidal orbifolds considered in this paper and the possible extensions suggested above are examples of global orbifolds, i.e.\ a geometrical construction $M^6/G$ where a six-dimensional manifold $M^6$ is divided by some finite group $G$. However, there also exist non-global orbifolds, where the orbifolding is performed patch-wise with certain compatibility requirements on overlaps (for details  see e.g.~\cite{Chen:2000cy}). For such non-global orbifolds, it might be possible that the orbifolding in each patch admits some Killing spinors, but there are no Killing spinors globally on all patches combined, leading to vanishing right-moving fermionic partition functions in a non-supersymmetric setting.


In addition, our non-existence result explicitly assumed, that we could make the partition function to vanish in all (twisted) sectors individually. As discussed in section~\ref{sec:vanishingcc} this is only the most conservative and model independent way one could achieve a vanishing (one-loop) cosmological constant (which did work in certain type II constructions). However, our results certainly do not exclude the possibility to have accidental cancellations between various sectors in the full partition function or that only the modular integral over the one-loop amplitude vanishes. 
It would be very interesting to investigate whether either option is at all possible and if so how to implement them in a systematic fashion.


Moreover, as stated above the mathematical conjecture~\ref{Conjecture} that underpinned our no-go result was only tested for a large but certainly not exhaustive number of finite groups. Therefore, a future direction for mathematical work could be to either prove this conjecture or to determine finite groups where it fails.


If it is impossible to have the cosmological constant to vanish at the one-loop level, maybe the 
next best thing is to have it be exponentially suppressed. Models with exponentially small 
cosmological constant at one loop, due to an accidental Bose-Fermi degeneracy only at the massless 
level, were constructed in~\cite{Abel:2015oxa} based on the idea of having heterotic constructions 
which are able to interpolate between supersymmetric and non-supersymmetric string 
vacua~\cite{Itoyama:1986ei,Dienes:1990ij,Dienes:1994np,Faraggi:2009xy}. In particular, heterotic 
models with small and positive cosmological constant have been constructed recently 
in~\cite{Florakis:2016ani} and a possible solution to the decompactification problem of such 
constructions was suggested e.g.\ in~\cite{Faraggi:2014eoa}. The results of our paper strengthen 
the motivation to consider such constructions to address the cosmological constant problem in 
non-supersymmetric heterotic string models.


Finally, in this paper we had to develop a couple of new techniques to systematically study non-Abelian orbifolds in order to understand the possible structures of the right-moving fermionic partition function. These techniques might prove very useful to investigate heterotic model building on non-Abelian supersymmetric or non-supersymmetric orbifolds in a more systematic fashion than was possible in the past (see e.g.~\ \cite{Konopka:2012gy,Fischer:2013qza}). 
In particular, the use of  representation theory of finite groups could provide a tool to efficiently characterize gauge bundles on non-Abelian orbifolds.

\subsubsection*{Acknowledgements}

We thank Carlo Angelantonj, Ignatios Antoniadis, Emilian Dudas and Andreas Trautner for useful 
discussions. This work was partially supported by the DFG cluster of excellence ``Origin and 
Structure of the Universe'' (www.universe-cluster.de) and by the Deutsche Forschungsgemeinschaft 
(SFB1258). The work of O.L. is supported by the Swiss National Science Foundation (\textsc{snf}) 
under grant number \textsc{pp}00\textsc{p}2\_157571/1.

\appendix
\def\theequation{\thesection.\arabic{equation}}

\section{Riemann Identities for Fermionic Partition Functions}
\label{App:AppRiemannId}
\setcounter{equation}{0}

In this appendix we present the structure of fermionic partition functions and discuss 
generalizations of the Riemann identity that could be used to determine when fermionic 
partition functions vanish.

\subsection{Fermionic Partition Functions}

In this appendix, we consider an even number of complex left-moving worldsheet fermions 
$\gl=(\gl^I)$, $I=1,\dots,d$, with the following boundary conditions 
\equ{
\gl^I(\gs+1) ~=~ e^{2\pi \I\, t/\gn} \, e^{2\pi \I\, \ga^I}\, \gl^I(\gs)~, 
\qquad 
\gl^I(\gs-\gt) ~=~ e^{2\pi \I\, t'/\gn} \, e^{2\pi \I\, \ga^{\prime\, I}}\, \gl^I(\gs)~. 
}
The integers $t,t' = 0, \ldots, \gn-1$ label the $\gn$-sectors of a $\Intr_\gn$-spin-structure. In particular, if we take $\gn=2$, this corresponds to the standard spin-structures of the $\text{Spin}(32)/\Intr_2$ or $\text{E}_8\times \text{E}_8$ heterotic string theories if we take $d=16$ or two bunches of $d=8$ fermions. The vectors $\ga, \ga' \in \Ratl^d$ parametrize further arbitrary boundary conditions that these left-moving fermions may fulfill.

We can define the following partition functions when summing over the spin-structures
\equ{ \label{LMpartitionFun} 
\mathcal{Z}_{d}\brkt{\ga}{\ga'}(\gt) ~=~ 
\frac 12\,e^{-\pi \I\, \ga^T(\ga'-e_d)}\,  \sum_{t,t'=0}^{\gn-1}\, e^{-\pi \I\, \sfrac d{\gn^2}\, t't}\, 
e^{-2\pi \I\, \sfrac{t'}\gn e_d^T \ga} \, 
 \frac{\gth_d\brkt{\sfrac12 e_d - \sfrac {t}{\gn} e_d- \ga}{\sfrac 12 e_d -\sfrac {t'}{\gn} e_d-\ga'}(\gt)}{\get^d(\gt)}~. 
}
Here $e_d = (1,\ldots, 1)^T$ is the $d$-component vector with all its entries equal to one. 
In addition, we use the $d$-dimensional theta function with $d$-component vector-valued 
characteristics $\gb$ and $\gb'$:
\equ{ \label{dDtheta}
\gth_d\brkt{\gb}{\gb'}(\gt) ~=~ 
 \sum_{n\in\Intr^d} e^{2\pi \I \big\{ \frac \gt2 (n+\gb)^2 + (n+\gb)^T\gb' \big\}}~. 
}
The partition functions~\eqref{LMpartitionFun} are modular covariant in the sense that 
\equ{ \label{ModCovTransPartition}
\hspace{-1ex} 
\mathcal{Z}_{d}\brkt{\ga}{\ga'}\big(\gt+1\big) ~=~ e^{2\pi \I \sfrac{d}{12}}\, \mathcal{Z}_{d}\brkt{\ga}{\ga'+\ga}\big(\gt\big)~, 
\qquad 
\mathcal{Z}_{d}\brkt{\ga}{\ga'}\big(\sfrac{-1}\gt\big) ~=~ 
e^{-2\pi \I \sfrac d4}\, 
\mathcal{Z}_{d}\brkt{\ga'}{-\ga}\big(\gt\big)~.
}
Hence, they can conveniently be used as building blocks for fully invariant partition functions.

\subsection[Generalized Riemann Identities for d-Dimensional Theta Functions]{Generalized Riemann Identities for \boldmath $d$-Dimensional Theta Functions}
\label{ssc:GeneralizedIdentities}


In the following, we do not take the order of the spin-structures $\gn$ and the number of complex fermions $d$ to be independent, but assume that they are related via $d=2\gn$. In this case we can derive various generalized Riemann identities, which may ensure that the corresponding partition functions vanish under certain conditions. These Riemann identities originate from a fractional symmetric orthogonal $d\times d$ matrix $S\in\text{SO}(d;\Ratl)$, i.e. 
\equ{
S^T ~=~ S~, 
\qquad  
S^T S ~=~ \Id_d\;.
}
Call $\widetilde{\ga} = S \ga$ the characteristic $\ga$ transformed by $S$. The Riemann identity 
then expresses the $d$-dimensional theta function with characteristics 
$(\widetilde{\ga}, \, \widetilde{\ga}')$ as a finite sum of the original ones over some new 
summation variables.


For example, we can consider the fractional matrix 
\equ{ \label{MatrixSEvenD}
S ~=~ \frac 1{\gn} \, e_d\, e_d^T - \Id_d~, 
}
which has the described properties above when $d = 2\gn$. 
(More generally, we could take $S = \frac 1{\gn} \, w\, w^T - \Id_d$ with a vector $w$ with all non-vanishing integers and $\gn = w^2/2$.) 
For any integral vector $n \in \Intr^d$, the vector 
\equ{ \label{SnTotm} 
\widetilde{n} ~=~ S\,n ~=~  \sfrac 1\gn\Big(\sum_i n_i\Big)\, e_d - n ~\stackrel{!}{=}~ \sfrac t\gn\, e_d + m~,
}
can be decomposed in an integral part, labeled by $m\in\Intr^d$, and a fractional part, labeled by $t = 0,\ldots, \gn-1$. (The latter distinguishes the $\gn$ sectors associated to a $\Intr_{\gn}$-spin structure.) However, not for any $t=0,\ldots \gn-1$ and $m\in \Intr^d$ the suggested identification of eq.~\eqref{SnTotm} make sense, since then we would be ignoring a constraint between $m$ and $t$. To identify this constraint, we first notice that eq.~\eqref{SnTotm} implies that $t = e_d^Tn - \gn \, \big[{e_d^Tn}/\gn\big]$ and $m =  \big[{e_d^Tn}/\gn\big] e_d -n$, where $[\ldots]$ restricts to the integral part. By taking the inner product with $e_d$ of the second equation, we uncover the constraint: 
\equ{ \label{Constraintmt} 
e_d^Tm + t \in \gn\, \Intr~. 
}
%


To obtain a generalized Riemann identity we write out 
$\gth_d\brkt{\widetilde{\ga}}{\widetilde{\ga}'}(\gt)$ using the identification~\eqref{SnTotm}
\equ{ 
\gth_d\brkt{\widetilde{\ga}}{\widetilde{\ga}'}\big(\gt) ~=~ 
\sum_{t =0}^{\gn-1} \sum_{m\in\Intr^d} 
\gd_{\gn\Intr}(e_d^Tm+t)\, 
e^{2\pi \I\big\{\sfrac \gt2\big(m+\sfrac t\gn e_d - \ga\big)^2 + \big(z-\ga'\big)^T\big(m+\sfrac t\gn e_d - \ga\big)  \big\}}~.  
}
where the constraint~\eqref{Constraintmt} is implemented using the order-$\gn$ delta function, which amounts to projecting out all summands that do not fulfill this constraint
\equ{ \label{DeltaZnu}
\gd_{\gn \Intr}(\gb) ~=~ \frac 1{\gn} \sum_{t'=0}^{\gn-1} e^{2\pi \I\, \sfrac {t'}\gn \gb}~. 
}
Inserting this expression, we can recognize the $d$-dimensional
theta functions and simplify it to 
\equ{ 
\gth_d\brkt{\widetilde{\ga}}{\widetilde{\ga}'}(\gt) ~=~ 
\frac 1{\gn}
\sum_{t',t =0}^{\gn-1}
\, e^{-2\pi \I\,\big\{ \sfrac{t't}\gn -\sfrac{t'}{\gn}\,e_d^T\ga\big\}} \, 
\gth_d\brkt{\ga-\sfrac{t}{\gn}\,e_d}{\ga'- \sfrac{t'}{\gn}\,e_d}(\gt)~. 
}
%
%
Since $S\, e_d = e_d$, we find that $\ga \mapsto \sfrac 12\, e_d - \ga$ is equivalent to $\widetilde{\ga} \mapsto \sfrac 12\, e_d - \widetilde{\ga}$. Making these substitutions we obtain the generalized 
Riemann identity 
\equ{ 
\frac 1{\gn}
\sum_{t',t =0}^{\gn-1}
\, e^{- 2\pi \I\, \big\{\sfrac{t't}\gn + \sfrac{t'}{\gn}\,e_d^T\ga\big\}} \, 
\gth_d\brkt{\sfrac 12\, e_d -\sfrac{t}{\gn}\,e_d- \ga}{\sfrac 12\, e_d- \sfrac{t'}{\gn}\,e_d -\ga'}(\gt)
~=~
\gth_d\brkt{\sfrac 12\, e_d -\widetilde{\ga}}{\sfrac 12\, e_d -\widetilde{\ga}'}(\gt)~. 
}
%


Now, notice that the left-hand-side appears in the form of a
fermionic partition function~\eqref{LMpartitionFun} with $\Intr_\gn$-spin structures. Hence, this partition function vanishes: 
\equ{ \label{VanishingPF}
\mathcal{Z}_d\brkt{\ga}{\ga'}(\gt) ~=~ 0~, 
}
if and only if the same entries of $\widetilde{\ga}$ and $\widetilde{\ga}'$ vanish modulo integers, since the expression 
then is proportional to $\gth\brkt{1/2}{1/2}$, which is zero.

\subsection{Right-moving Fermionic Partition Function}
\label{app:RMPartitionFunction}


The partition function of the right-moving fermions $\gps_\text{R}=(\gps_\text{R}^a)$, used in the main text, is most conveniently determined as the complex conjugate of the partition function of left-moving fermions with the same boundary conditions. 
Their orbifold boundary conditions are determined by the local twist vectors $v_g=\big(0,v_g^1,v_g^2,v_g^3\big)$ associated to a given space group element $g\in \mathbf{P}$. (The first zero in this vector just means that we have trivial boundary conditions in the four-dimensional Minkowski space corresponding to the fermion $\gps_\text{R}^0$.) In addition, these fermions are defined with a $\Intr_2$-spin structure, hence their partition function reads
\equ{  \label{RMpartitionFunctionOfPsi} 
\mathcal{Z}_\gps\brkt{g}{h}(\tau) ~=~ \mathcal{Z}_4\brkt{v_g}{v_{h}}(\gt)~, 
}
using eq.~\eqref{ModCovTransPartition} of appendix~\ref{App:AppRiemannId} with $\gn=2$. 


This shows that we may apply the vanishing result~\eqref{VanishingPF} derived above for $d=4$ and $\gn=2$. Hence, we see that 
\equ{ \label{tildeLocalTwist}
\widetilde{v}_g ~=~ S\, v_g ~=~ 
\frac 12 \pmtrx{
\phantom{-}v_g^1 + v_g^2 +  v_g^3 \\[1ex] 
-v_g^1 + v_g^2 +  v_g^3 \\[1ex] 
\phantom{-}v_g^1 - v_g^2 +  v_g^3 \\[1ex] 
\phantom{-}v_g^1 + v_g^2 -  v_g^3 
}
\quad\text{where}\quad 
 S ~=~ \frac 12 \pmtrx {
-1 & \phantom{-}1 & \phantom{-}1 & \phantom{-}1 \\[1ex] 
\phantom{-}1 & -1 & \phantom{-}1 & \phantom{-}1 \\[1ex]
\phantom{-}1 & \phantom{-}1 & -1 & \phantom{-}1 \\[1ex] 
\phantom{-}1 & \phantom{-}1 & \phantom{-}1 & -1 
}~, 
}
precisely corresponds to the four eigenvalues of $D_\text{s}(\gth)$. Hence, as stated in the main text, the partition function $\mathcal{Z}_\gps\brkt{g}{h}$ vanishes iff the same components of $\widetilde{v}_g$ and $\widetilde{v}_{h}$ vanish modulo integers at the same time, i.e. $v_g$ and $v_{h}$ preserve the same Killing spinor. 


The choice of the matrix $S$ is not unique, one could also have started with the matrix 
\equ{ \label{MatrixS4D}
S = \frac 12 \left(\arry{rrrr}{ 
1 & 1 & 1 & 1 \\[1ex] 
1 & 1 & -1 & -1 \\[1ex] 
1 & -1 & 1 & -1 \\[1ex] 
1 & -1 & -1 & 1
}\right)~. 
}
The Riemann identity in this case will look somewhat different, but in the end one arrives at the same physical conditions as to when the right-moving fermionic partition function vanishes.

\subsection{Left-Moving Fermionic Partition Functions} 
\label{app:LMfermiPartition} 


As we have seen in this appendix, Riemann identities can naturally be applied to certain fermionic 
partition functions. Therefore, in order to investigate options to apply generalized 
Riemann identities to the left-moving gauge degrees of freedom of the heterotic string, 
it is most convenient to work in the fermionic formulation. In that case one has 16 left-moving 
complex fermions. In the standard $\text{Spin}(32)/\Intr_2$ or $\text{E}_8\times\text{E}_8$ 
theories there are one or two $\Intr_2$-spin structures, respectively. 


Since all Riemann identities under investigation have a spin structure of order 
$\gn = w^2/2 \geq d/2$, we realize that it is impossible to have Riemann identities for bunches of 
8 or 16 fermions with a $\Intr_2$-spin structure. Hence, the standard left-moving partition 
functions do not vanish. If one would allow for more exotic theories with $\Intr_4$- or 
$\Intr_8$-spin structures, then our results can provide Riemann identities and possible vanishing 
partition functions. Moreover, if one introduce four $\Intr_2$-spin structures for four bunches of 
four left-moving fermions, then the results given here can be applied. (Combinations of these 
options could also be considered.) However, in all of these cases the trivial vacuum state has been 
projected out, such that the remaining states can organize themselves in such a way that 
cancellations at all mass levels are possible. As indicated in the main text, this is not 
acceptable for heterotic string phenomenology since this means in particular that the  {\em massless} graviton will be absent.

\section{Some Representation Theory of SO(6), Spin(6) and SU(4)}
\label{App:Spin} 
\setcounter{equation}{0}

\subsection{SO(6): The Vector Representation} 

The generators $J_{ij}$ of the $\text{so}(6)$ Lie-algebra in the vector representation can be 
written as 
\equ{\label{eq:SO6VectorGenerators}
(J_{ij})_{kl} ~=~ \gd_{ik}\gd_{jl}-\gd_{jk}\gd_{il} 
}
for $i,j=1,\ldots,6$.
The $(i,j)$ and $(j,i)$ entry of the matrix $J_{ij}$ equals $+1$ and $-1$, respectively. 
Hence, $J_{ij}$ for $i<j$ and $i,j = 1,\ldots, 6$ form a basis of anti-symmetric $6 \times 6$ 
matrices. $J_{ij}$ generates a rotation in the $(X^i,X^j)$-plane. Then, a general $\SO{6}$ group 
element $D_\rep{v}(\gth)$ can be obtained by exponentiating the $\text{so}(6)$ anti-symmetric 
Lie-algebra parameter $\go_{ij}$ as 
\equ{ \label{Def:VectorRep} 
D_\rep{v}(\gth) ~=~ \exp \Big( \sfrac 12\, \go_{ij} \, J_{ij} \Big)~,
}
where $\gth\in \mathbf{P}$ denotes the corresponding abstract point group element and the factor 
$\frac{1}{2}$ in eq.~\ref{Def:VectorRep} accounts for the summation over the full index range of 
$i,j$ using the anti-symmetry of $\go_{ij}$ and $J_{ij}$.

\subsection{Spin(6): The Spinor Representation} 
\label{App:Spin_SpinorRep} 

The six-dimensional Euclidean Clifford algebra is generated by $8\times 8$ Hermitian 
gamma matrices $\gG_i$, $i=1,\ldots,6$, and the chirality operator 
$\tgG = \I \, \gG_{1} \, \gG_{2} \dots\gG_{6}$, which satisfy
\equ{ \label{CliffordAlgebra} 
\big\{ \gG_i, \gG_j \big\} ~=~ 2\, \gd_{ij}\, \Id_8~, 
\qquad 
\tgG^2 ~=~\Id_{8}~, 
\qquad 
\big\{ \gG_i, \tgG \big\} ~=~ 0~.  
}
The chirality operator allows us to define the Weyl-spinors with help of the projectors,
\begin{equation}\label{eq:weylprojector}
P^{(\pm)}~=~ \dfrac{\Id_8 \pm \tgG}{2} 
\quad\text{with}\quad
 \Big(P^{(\pm)}\Big)^2 ~=~ P^{(\pm)}   
 \quad\text{and}\quad 
 P^{(+)}\, P^{(-)} ~=~ 0~.
\end{equation}
In addition, the charge conjugation matrix $C$ has the properties 
\equ{ \label{ChargeConjugation} 
C \,\gG_i\, C^{-1} ~=~ \gG_i^T~,
\qquad 
C \,\tgG\, C^{-1} ~=~ -\tgG^T~,
\qquad 
C^\dag ~=~ - C^T ~=~ C~. 
}
Consequently, $C\, P^{(\pm)}\, C^{-1} = (P^{(\mp)})^T$ and $(C\gG_i)^T = - C\gG_i$ are anti-symmetric matrices.

The spin representation is defined as 
\begin{align}\label{Def:SpinorRep}
D_\rep{s}(\gth) ~=~ \exp\Big(\sfrac{1}{2} \, \go_{ij} \; \gS_{ij} \Big) 
\quad\text{with}\quad 
 C \,D_\rep{s}(\gth) \,C^{-1} ~=~ D_\rep{s}(\gth^{-1})^T = D_\rep{s}(\gth)^*~,
\end{align}
in terms of the anti-Hermitian spin generators $\gS_{ij}$, satisfying
\begin{align}\label{eq:SpinGenerators}
\gS_{ij} ~=~ -\gS_{ji} ~=~  \sfrac{1}{4} \big[ \gG_{i} , \gG_{j}\big]~, 
\qquad 
C \,\gS_{ij}\, C^{-1} ~=~ - \gS_{ij}^T~,
\quad\text{and}\quad 
\text{Tr}(\gS_{ij}) = 0~.
\end{align}
These generators $\gS_{ij}$ fulfill the $\text{so}(6)$ Lie-algebra that can be defined from 
the generators $J_{ij}$ of the vector representation of $\text{so}(6)$ given in 
eq.~\eqref{eq:SO6VectorGenerators}. Furthermore, $D_\rep{s}(\gth)$ is a unitary matrix, since $D_\rep{s}(\gth)^\dag = D_\rep{s}(\gth^{-1}) = D_\rep{s}(\gth)^{-1}$. The 
eight-dimensional spin representation $D_\rep{s}$ acts naturally on eight-component Dirac spinors. 
The group generated by these spin generators is called the spin group $\text{Spin}(6)$.

%
The relation between the vector and spinor representation is given by 
\equ{  \label{DoubleCover} 
D_\rep{s}(\gth)^{T} \,C\gG_i\, D_\rep{s}(\gth) ~=~ \big[D_\rep{v}(\gth)\big]_{ij} \, C\gG_j~. 
}
This shows that the spin group $\text{Spin}(6)$ is the double cover of the special orthogonal group 
$\SO{6}$: the spin group elements $-D_\rep{s}(\gth)$ and $D_\rep{s}(\gth)$ both project to the 
same $\SO{6}$ group element $D_\rep{v}(\gth)$.  This means that if we are given a set of the 
generators $\{D_\rep{v}(\gth_\ga)\}$ of the geometrical point group $\mathbf{P}_\rep{v} \subset \SO{6}$, 
then for each of them we have two choices to define the corresponding $\text{Spin}(6)$ generators: 
$D_\rep{s}(\gth_\ga)$ or $-D_\rep{s}(\gth_\ga)$. On the other hand, $D_\rep{s}(\gth)$ determines 
$D_\rep{v}(\gth)$ uniquely, i.e. 
\equ{ \label{UniqueSOrep} 
\big[D_\rep{v}(\gth)\big]_{ij} ~=~ \sfrac 18\, \text{Tr} \big[ D_\rep{s}(\gth^{-1}) \,\gG_i\, D_\rep{s}(\gth) \,\gG_j \big]~, 
}
using eq.~\eqref{DoubleCover} and eq.~\eqref{Def:SpinorRep}. 
Because $D_\rep{v}(\gth)_{ji} = D_\rep{v}(\gth^{-1})_{ij} = \left(D_\rep{v}(\gth)^{-1}\right)_{ij}$, 
the matrix $D_\rep{v}(\gth)$ is orthogonal.  
Furthermore, $D_\rep{v}(\gth)$ is real, i.e.\ 
$D_\rep{v}(\gth)_{ij}^* = D_\rep{v}(\gth)_{ij}$ using $\gG_i^* = \gG_i^T$ and 
$D_\rep{s}(\gth)^* = D_\rep{s}(\gth^{-1})^T$. Hence, $D_\rep{v}(\gth)$ as defined in 
eq.~\eqref{UniqueSOrep} is an element of $\SO{6}$.


For the Clifford algebra the following variant of Schur's lemma holds: 
Let $M$ be a Clifford algebra matrix satisfying 
\equ{ \label{CliffordSchurLemma} 
C\, M\, C^{-1} ~=~ M^{-T}
\quad\text{and}\quad 
M^T\, C\gG_i\, M ~=~ C\gG_i~, 
}
for all $i=1,\ldots,6$, then $M=\pm \Id_8$. Indeed, these two equations can be combined to the condition $[M,\gG_i]=0$ for all $i$. Since the six-dimensional Clifford algebra is spanned by the 64 matrices, 
%
$\Id_8~,~ \gG_i~,~ \gG_{ij}~,~ \gG_{ijk}~,~ \gG_{ijk}\tgG~,~ \gG_{ij}\tgG~,~ \gG_i\tgG~,~ \tgG$ 
%
(where $\gG_{ij}$ and $\gG_{ijk}$ are the two and three times totally anti-symmetric products of gamma matrices, respectively), we infer that only when $M = a\, \Id_8$ with $a\in\Cplx$, this matrix commutes with all gamma matrices.\footnote{This conclusion can also be reached by realizing that $\gG_i$ generate an irreducible representation of a  finite group of order 128 (GAP ID [128, 2327]) 
and applying Schur's lemma for finite groups.}
Inserting this back into the second relation of eq.~\eqref{CliffordSchurLemma}, we find $a^2=1$. 
Identifying $M$ with $D_{s}(\theta_{1})\cdots D_{s}(\theta_{n})$ and use $D_{\text{v}}(\theta_{1})\cdots D_{\text{v}}(\theta_{n})=\Id$, we thus obtained eq.~\eqref{ModificationsDefRelations}.

\subsection{SU(4):  Chiral Spinor Representations}
\label{ChiralBasis} 


To obtain a convenient chiral basis of the six-dimensional Clifford algebra, we express the gamma matrices $\gG_i$, chirality operator $\tgG$ and charge conjugation $C$ as 
\equ{\label{6DgammasChiral} 
\gG_i ~=~ \pmtrx{ 0 & \bgg_i \\ \gg_i & 0}~, 
\quad 
\gS_{ij} ~=~ \pmtrx{ \gs_{ij} & 0 \\ 0 &  \bgs_{ij}}~, 
\quad 
\tgG ~=~ \pmtrx{ \Id_4 & 0 \\ 0 & - \Id_4 }~, 
\quad 
C ~=~ \pmtrx{0 & c \\ c & 0}~, 
}
in terms of Clifford algebra generators $\bgg_\gk = \gg_\gk$, $\gk=1,\ldots, 5$, and charge conjugate matrix $c$, satisfying $c \,\gg_\gk\, c^{-1} = \gg_\gk^T$ and $c^\dag = c^{-1} = -c^T$  in five dimensions and $-\bgg_6 = \gg_6 = \I \, \Id_{4}$. The spin generators~\eqref{eq:SpinGenerators} are expressed in terms of (two sets of 15) anti-Hermitean $\text{su}(4)$ generators,  
\equ{
\gs_{ij} ~=~ \sfrac 14 ( \bgg_i \gg_j - \bgg_j \gg_i)
\quad\text{and}\quad  
\bgs_{ij} ~=~ \sfrac 14 ( \gg_i \bgg_j - \gg_j \bgg_i)~, 
}
that are all traceless and related to each other via 
$\bgs_{ij} = - c^{-1} \, \gs_{ij}^T\, c$.


The spin representation~\eqref{Def:SpinorRep} is a reducible representation.
Using the projectors~\eqref{eq:weylprojector} we can define the two irreducible chiral representations 
$D^{(\pm)}_\rep{s}(\gth)  = P^{(\pm)}D_\rep{s}(\gth)$, which can be identified with the $\SU{4}$-matrices
\equ{
D_\rep4(\gth) ~=~ \exp\Big(\sfrac{1}{2} \, \go_{ij} \; \gs_{ij} \Big)  
\qquad  
D_{\crep{4}}(\gth) ~=~ \exp\Big(\sfrac{1}{2} \, \go_{ij} \; \bgs_{ij} \Big)~, 
}
generated by the matrices $\gs_{ij}$ and $\bgs_{ij}$, as
\equ{ \label{ReductionChiralBasis}
D^{(+)}_\rep{s}(\gth) ~=~ \pmtrx{ D_\rep4(\gth) & 0 \\ 0 & 0 }~,
\qquad 
D^{(-)}_\rep{s}(\gth) ~=~ \pmtrx{ 0 & 0 \\ 0 & D_{\rep{\bar{4}}}(\gth) }~, 
}
respectively. Using the anti-Hermiticity and tracelessness of $\gs_{ij}$, it is not difficult to confirm, that $D_\rep{4}(\gth)^\dag = D_\rep{4}(\gth^{-1}) = D_\rep{4}(\gth)^{-1}$ is an $\SU{4}$ element. Moreover, the representations, $\rep{4}$ and ${\crep{4}}$, are related via charge conjugation~\eqref{Def:SpinorRep} as $D_{\crep{4}}(\gth) = c^{-1}\,D_\rep{4}(\gth^{-1})^{T}\, c$.


An explicit realization of the representation  $\rep{6}=[\rep{4}]_2$ of $\SU{4}$ is given by anti-symmetric $4\times 4$-matrices , $A^T=-A$, transforming as 
\equ{
A ~\mapsto~ D_\rep{4}(\gth)^T A\, D_\rep{4}(\gth)
}
under the action of $D_\rep{4}(\gth)\in\SU{4}$. Any set of six linear independent anti-symmetric $4\times 4$-matrices, forms a basis for $\rep{6}$. Since the anti-symmetric matrices $C\gG_i$ are given by 
\equ{ 
C\gG_i ~=~ \pmtrx{ c\gg_i & 0 \\ 0 & c\bgg_i}
}
in the basis~\eqref{6DgammasChiral}, we infer that $c\gg_i$ define six different anti-symmetric $4\times 4$-matrices. Hence, we can expand $A= A_i\, c\gg_i$. The $A_i$ transform as the six components  of the $\SO{6}$-vector: $A_i \mapsto [D_\rep{v}(\gth)]_{ij}\, A_j$, since eq.~\eqref{DoubleCover} 
implies that $D_\rep{4}(\gth)^T c\gg_i\, D_\rep{4}(\gth) = [D_\rep{v}(\gth)]_{ij}\, c\gg_j$.

\subsection{An Explicit Spin(6) Basis}
\label{SecSpin6Basis}

A convenient choice for the $\gG$-matrices in six dimensions is given by
\equ{\label{6Dgammas} 
\arry{l}{
\gG_1 ~=~ \gs_1 \otimes \,\Id_2\, \otimes \,\Id_2~, 
\\
\gG_2 ~=~ \gs_2 \otimes \,\Id_2\, \otimes \,\Id_2~, 
\\ 
}
\quad
\arry{l}{
\gG_3 ~=~ \gs_3 \otimes \gs_1 \otimes \, \Id_2~, 
\\
\gG_4 ~=~ \gs_3 \otimes \gs_2 \otimes \, \Id_2~,
}
\qquad 
\arry{l}{
\gG_5 ~=~ \gs_3 \otimes \gs_3 \otimes \gs_1 ~, 
\\
\gG_6 ~=~ \gs_3 \otimes \gs_3 \otimes \gs_2 ~. 
}
}
One can check that with this definition the Clifford algebra~\eqref{CliffordAlgebra} is fulfilled. 
In this basis, the chirality operator reads
\equ{ 
\tgG  ~=~  \gs_3 \otimes \gs_3 \otimes \gs_3~.
}
Moreover, using this basis the corresponding Cartan algebra of $\text{Spin}(6)$ can be represented 
by commuting spin generators $\gS_{1} = \gS_{12}$, $\gS_2 = \gS_{34}$ and $\gS_3 = \gS_{56}$ given 
by 
\equ{ \label{SigmaBasis}
\gS_{1} ~=~ \sfrac{\I}2\, \gs_3 \otimes \,\Id_2\, \otimes \,\Id_2~, 
\quad 
\gS_2  ~=~ \sfrac{\I}2\,\, \Id_2\, \otimes \gs_3 \otimes \,\Id_2~, 
\quad
\gS_3  ~=~ \sfrac{\I}2\,\, \Id_2\,  \otimes \,\Id_2\, \otimes \gs_3~. 
}
%


\section{Elements of Finite Group Theory}
\label{app:FiniteGroup} 
\setcounter{equation}{0}

In this appendix we collect a number of facts about finite groups and their 
representations. Details can be found for example in~\cite{Isaacs:1994ch}.


Let $\mathbf{G}$ denote a finite group of order $|\mathbf{G}|$, where $|\mathbf{G}|$ is the number 
of elements in $\mathbf{G}$. The elements of $\mathbf{G}$ can be divided into disjoint conjugacy 
classes  $[\gth_x] = \big\{ \gr \,\gth_x \gr^{-1} \big| \forall \gr \in \mathbf{G} \big\}$, where 
$x=1,\ldots, c$. Here, $c$ denotes the number of conjugacy classes and $\gth_x\in\mathbf{G}$ is a 
representative of the set $[\gth_x]$. The unit element $\Id$ always forms its own conjugacy class 
$[\Id]=\big\{\Id\big\}$. We define an ordered tuple of all conjugacy classes 
$([\gth_1],\ldots,[\gth_c])$ in which $[\gth_1] = [\Id]$ appears in the first place.


A representation $\rep{R}$ defines a group homomorphism 
$D_{\rep{R}}: \mathbf{G} \ra \text{GL}(|\rep{R}|; \Cplx)$, where $|\rep{R}|$ is the dimension (also 
called degree) of the representation, i.e.\ $|\rep{R}|$ is the dimension of the vector space the 
representation $\rep{R}$ acts on. In fact, all representations of any finite group can be chosen to 
be unitary, i.e.\ $D_\mathbf{R}(\gth) \in U(|\mathbf{R}|)$ for all $\gth\in\mathbf{G}$. New 
representations can be built from some given representations by taking direct sums $\oplus$ and 
tensor products $\otimes$. In particular, the $k$ times anti-symmetrized tensor product of the 
representation $\rep{R}$ is denoted by $[\rep{R}]_k$.


To each representation $\rep{R}$ one can associate a character 
$\gch_\rep{R}: \mathbf{G} \ra \Cplx$  
\equ{ \label{DefCharacter} 
\ch{\rep{R}}{\gth} ~=~ \text{Tr}(D_\rep{R}(\gth))~.  
}
In particular, the dimension of $\rep{R}$ can be determined by $|\rep{R}| = \ch{\rep{R}}{\Id}$. 
Characters are class functions in the sense that 
$\ch{\rep{R}}{\gth} = \ch{\rep{R}}{\gr \,\gth_x \gr^{-1}}$ for all $\gr\in\mathbf{G}$, which we 
thus denote by $\ch{\rep{R}}{[\gth]}$. A representation $\rep{R}$ is uniquely identified by the 
$c$-component vector
\equ{ \label{ChValues} 
\gch_\rep{R} ~=~ 
\Big( \ch{\rep{R}}{[\gth_1]}, \ldots, \ch{\rep{R}}{[\gth_c]}\Big)
}
of its characters evaluated for all the conjugacy classes. 

Characters of direct sums and tensor products equal sums and products of the corresponding characters 
\equ{ \label{SumProdChar} 
\ch{\rep{R}\oplus \rep{R}'}{\gth} ~=~ \ch{\rep{R}}{\gth} + \ch{\rep{R}'}{\gth}~, 
\qquad 
\ch{\rep{R}\otimes \rep{R}'}{\gth} ~=~ \ch{\rep{R}}{\gth} \, \ch{\rep{R}'}{\gth}~.
}
This can be used to relate the characters of $k$ times anti-symmetrized representations $[\rep{R}]_k$ 
to the characters of the original representation $\rep{R}$. In particular, we have that~\cite{vanRitbergen:1998pn,GrootNibbelink:2003gb}
\equa{ \label{AntiSymCharacters}
\ch{[\rep{R}]_2}{\gth} &=~ \frac{1}{2}\left(\left(\ch{\rep{R}}{\gth}\right)^2 - \ch{\rep{R}}{\gth^2}\right)~, 
\\[1ex] 
\ch{[\rep{R}]_4}{\gth} &=~ 
 \frac{1}{4!}\left(
 \left(\ch{\rep{R}}{\gth}\right)^4 
 - 6\, \ch{\rep{R}}{\gth^2} \,  \left(\ch{\rep{R}}{\gth}\right)^2
+ 8\,  \ch{\rep{R}}{\gth^3} \,  \ch{\rep{R}}{\gth}
+3\, \left(\ch{\rep{R}}{\gth^2}\right)^2 
- 6\, \ch{\rep{R}}{\gth^4}
 \right)~.
 \non 
}
On the space of characters one can define the following inner product 
\equ{ \label{ChInner} 
\langle \gch_\rep{R}, \gch_{\rep{R}'}\rangle_{\mathbf{G}} ~=~ \frac{1}{|\mathbf{G}|}
\sum_{\gth' \,\in\, \mathbf{G}} \overline{\ch{\rep{R}}{\gth'}} \, \ch{\rep{R}'}{\gth'}~, 
}
for any two characters associated to representations $\rep{R}$ and $\rep{R}'$ of $\mathbf{G}$.


We denote the irreducible representations of $\mathbf{G}$ with lower case letters as $\rep{r}_x$. 
In particular, $\rep{1}$ refers to the trivial one-dimensional representation of $\mathbf{G}$ (also 
called singlet representation), where $D_{\rep{1}}(\gth) = 1$ for all $\gth \in \mathbf{G}$ and, 
hence, $\gch_\rep{1} = (1,\ldots, 1)$. The number of irreducible representations $\rep{r}_x$ of a 
finite group $\mathbf{G}$ is finite and equal to the number $c$ of conjugacy classes. We assume 
that the tuple of irreducible representations $(\rep{r}_1,\ldots,\rep{r}_c)$ is partially 
ordered by increasing dimension of the representations and the trivial singlet representation 
$\rep{r}_1 =\rep{1}$ appears first. 


The characters of irreducible representations are orthonormal,
\equ{ \label{OrthoCharacters}
\langle \gch_{\rep{r}_x}, \gch_{\rep{r}_y}\rangle_{\mathbf{G}} ~=~ \gd_{xy}~, 
}
w.r.t.\ the inner product~\eqref{ChInner}. The $c \times c$\,-\,matrix 
$T\in \text{GL}(c\,; \Cplx)$ of the so-called character table $T$ is defined by
\equ{ \label{MatrixCharacterTable}
T_{xy} ~=~ \ch{\rep{r}_x}{[\gth_y]}~.
}
%


Any representation $\rep{R}$ can be decomposed into irreducible representations 
\equ{  \label{IrrepDecom} 
\rep{R}  ~=~ 
\bigoplus_{x=1}^c  \left(\rep{r}_x\oplus\right)^{n_x}~, 
}
where $\left(\rep{r}_x\oplus\right)^{n_x} = \rep{r}_x \oplus \cdots \oplus \rep{r}_x$ ($n_x$ times) 
and the non-negative coefficient $n_x\in \mathbbm{N}_0$ counts how often the irreducible 
representation $\rep{r}_x$ lies inside $\rep{R}$. Using eqns.~\eqref{SumProdChar} and the 
orthogonality of the irreducible characters~\eqref{OrthoCharacters} we find that 
\equ{ \label{IrrepCharDecom} 
\ch{\rep{R}}{\gth} ~=~ \sum_{x=1}^c n_x\, \ch{\rep{r}_x}{\gth}~, 
\quad \text{with} \quad 
n_x ~=~ \langle \gch_{\rep{r}_x}, \gch_\rep{R} \rangle_{\mathbf{G}}~. 
}
The coefficients $n_x$ can also be computed directly using the matrix of the character 
table~\eqref{MatrixCharacterTable}, i.e.
\equ{ \label{IrrepDecomCoeffs} 
n_\rep{R} ~=~ \gch_\rep{R} \, T^{-1}~,
}
where $n_\rep{R} = (n_1, \ldots, n_c)$.


\subsection*{Branching to \boldmath $\Z{N}$ representations}

Since all elements of the cyclic group $\Z{N}$ commute, all $N$ group elements define their own 
conjugacy classes. Hence, there are are also $N$ irreducible representations of $\Z{N}$; all 
being one-dimensional. They are denoted by $\rep{1}_q$ for $q=0,\ldots, N-1$ and $\rep{1}_0$ labels 
the trivial representation. Their characters are equal to the representation itself, i.e.
\equ{ \label{ZNirreprs}
\gch_{\rep{1}_q}(\gth) ~=~
D_{\rep{1}_q}(\gth) ~=~ \exp\left(2\pi\I q/N\right)~,
}
for the $\Z{N}$ generator $\gth$ (with $\gth^N = \Id$).

Now, let $\Z{N}$ be a subgroup of $\mathbf{G}$ generated by an order $N$ element $\gth \in \mathbf{G}$.
To find the branching of the representation $\rep R$ of the full group $\mathbf{G}$ into the irreducible representations of a $\Z{N}$-subgroup, we use the character inner product~\eqref{ChInner} for the subgroup $\Z{N}$:
\equ{ \label{ZNbranching} 
n_{\mathbf{R};q} ~=~ 
\langle \gch_\rep{R}, \gch_{\rep{1}_q}\rangle_{\Z{N}} ~=~ \frac{1}{N}
\sum_{\gth' \,\in\, {\Z{N}}} \overline{\ch{\rep{R}}{\gth'}} \, \ch{\rep{1}_q}{\gth'}~, 
}
gives the number of times the irreducible representation $\rep{1}_{q}$ appears in the branching. Note that in this inner product we used characters defined w.r.t.\ two different groups, i.e.\ 
$\gc_{\rep R}$ and $\gc_{\rep{1}_q}$ for $\mathbf{G}$ and its $\Intr_N$-subgroup, respectively.

\section[Four-Dimensional Representation D4 with det(D4) != 1]{Four-Dimensional Representation \boldmath $D_\rep{4}$ with \boldmath $\det(D_\rep{4})\neq 1$}
\label{app:T7Example}

In order to highlight the assumptions behind our conjecture~\ref{Conjecture} we give here 
a second extension (see section~\ref{sec:ViolatingConjecture} for the first extension): we loosen 
the assumption $\det(D_\rep{4})= 1$ and easily find counter examples to this extended conjecture.

Consider the finite group $T_7$ of order 21, see e.g.~\cite{Ishimori:2010au}. $T_7$ has five 
irreducible representations: a three-dimensional representation $\rep3$ and its complex conjugate 
$\crep3$, as well as the trivial singlet $\rep1_0$ plus two non-trivial singlets $\rep1_1,\rep1_2$. 
Furthermore, $T_7$ has two different $\Z{N}$ subgroups: $\Z3$ and $\Z7$. The one-dimensional 
representation $\rep1_q$ ($q=0,1,2$) of $T_7$ branches to $\rep1_q$ of $\Z{3}$ and to $\rep1_0$ of 
$\Z{7}$, respectively. Let us choose the four-dimensional representation $\rep4$ of $T_7$ to be
\begin{align}
\rep4 ~=~ \rep3 \oplus \rep1_1\;.
\end{align}
This representation clearly does not contain a trivial singlet. Hence, condition {\em ii.} of the 
extended conjecture~\ref{Conjecture} is fulfilled. Using the branching of the various representations 
into representations of the subgroups we see that for the $\Z3$ subgroup
\begin{equation}\label{eq:4ofT7Z3branching}
\raisebox{9.2mm}{$\rep{4}~=~$}
\renewcommand{\arraystretch}{1.3}
\begin{array}{clcc}
   \rep{3}                                                    & \oplus \phantom{M}& \rep{1}_1   & \quad\text{of }T_7 \\
         \downarrow                                                 &        & \downarrow & \\ 
            \rep{1}_0 \oplus \rep{1}_1 \oplus \rep{1}_2 & \oplus & \rep{1}_1  & \quad\text{of }\Z3
        \\ \hline
            \rep{1}_1 \oplus \rep{1}_2 \oplus \rep{1}_4 & \oplus & \rep{1}_0  & \quad\text{of }\Z7
\end{array}
\end{equation}
Therefore, this representation also fulfills condition {\em iii.} of the extended conjecture~\ref{Conjecture}.
It has exactly the branching behaviour we were looking for: for each $\Z{N}$ subgroup of $T_7$ 
there exists a trivial singlet of $\Z{N}$.

%

%
%

\providecommand{\bysame}{\leavevmode\hbox to3em{\hrulefill}\thinspace}
\frenchspacing
\newcommand{\origttfamily}{}
\let\origttfamily=\ttfamily
\renewcommand{\ttfamily}{\origttfamily \hyphenchar\font=`\-}

\end{document}